\documentclass[12pt,a4paper,oneside,final]{book}

\usepackage{listings}
\usepackage{enumitem}

\usepackage[utf8]{inputenc} 
\usepackage{hyphenat}
\usepackage{graphicx}
\usepackage{graphicx}
\usepackage{amsmath,amssymb,amsthm}
\usepackage{caption}
\usepackage{multicol}
\usepackage[section]{placeins}
\usepackage{float}
\usepackage{subcaption}
\usepackage{mathptmx}
\usepackage{longtable,lipsum}

\usepackage[english]{babel}

\usepackage[Mgr,Tisk]{sci.muni.thesis}

\NazevUstavu{Ústav teoretické fyziky a~astrofyziky}{Department of Theoretical Physics and Astrophysics}

\RokOdevzdaniPrace{2016}

\AkademickyRok{2015/2016}

\Autor{Romana Grossová}{Bc. Romana Grossová}

\NazevPrace{Vysokodisperzní~\

spektroskopie~s
~Ondřejovským~Ešeletovým
~Spektrografem}{Vysokodisperzní~spektroskopie~
s~Ondřejovským~ 
Ešeletovým Spektrografem}{High Dispersion Spectroscopy with Ondřejov Echelle Spectrograph}

\VedouciPraceSTituly{RNDr. Petr Škoda, CSc.} 

\StudijniProgram{Fyzika}{Physics}

\StudijniObor{Teoretická fyzika a~astrofyzika}{Theoretical physics and astrophysics}

\PocetStran{vii\,$+$\,71}

\KlicovaSlova{Ešeletový spektrograf, IRAF, OPERA, exoplanety}{Echelle spectrograph, IRAF, OPERA, exoplanets}

\Abstrakty%
{Ešeletové spektrografy a jejich vysoké rozlišení hrají důležitou roli při určování\
cha-\
rakteristik hvězdných čar. Rozsáhlé pole aplikací je zaměřeno především na měření přes-\
ných radiálních rychlostí. V této diplomové práci se věnujeme kalibraci Ondřejovského ešeletového spektrografu v Astronomickém ústavu Akademie věd České Republiky. Mým úkolem bylo prozkoumat široké pole možností, jak zpracovávat data s dosažením nej\-lep\-ších možných výsledků. Úspěšná redukce byla provedena jak pro Image Reduction and Analysis Facility (IRAF) tak pro Open source Pipeline for ESPaDOnS Reduction and Analysis (OPERA). Tato diplomová práce zahrnuje srovnání obou redukčních systémů.}%
{Echelle spectrographs with their high resolution plays important role in determination of characteristics of stellar lines. Wide field of applications is focused mainly on the measurements of precise radial velocity applied in exoplanetary research. In my diploma thesis I~am concentrated on the calibration of the Ondřejov echelle spectrograf at Astronomical Institute of the Czech Academy of Sciences. My role was to investigate the wide field of opportunities how to process the data with the best possible results. Successful reduction was performed by both Image Reduction and Analysis Facility (IRAF) and for Open source Pipeline for ESPaDOnS Reduction and Analysis. This thesis includes the comparison of both pipelines.\\
}

\TextPodekovani%
{Na tomto místě bych chtěla poděkovat svému vedoucímu Petrovi Škodovi za ochotu, pomoc, vedení a konzultace. Jeho kolegovi Miroslavovi Šlechtovi za poskytnutí dat, Zdeň-\
kovi Janákovi za jeho šikovné řešení důležitých problémů, Ederovi Martiolimu za jeho pozitívní přístup, návody, užitečné rady a mailové konzultace, Samuelovi Molnárovi za informatickou a morální podporu a v neposlední řadě i své rodině a přátelům za podporu.\\

I would like to thank my supervisor Petr Škoda for his help, leadership and consultations and his colleague Miroslav Šlechta for providing data. To Zdeněk Janák for his clever solutions of important issues. Big thanks goes to Eder Martioli for his possitive attitude and leadership, instructions, useful advices and mail consultations. To Samuel Molnar for his technical and moral support and last but not least to my family and friends for their support.   }

\TextProhlaseni%
{Prohlašuji, že jsem svoji diplomovou práci vypracovala samostatně
 s využitím informačních
zdrojů, které jsou v~práci citovány.}

\DatumProhlaseni{5. ledna 2015}

\makeindex

\begin{document}

\VytvorPovinneStrany

\AbstraktyNaJedneStrane

\newpage
\PodekovaniAProhlaseni

\pdfbookmark{Obsah}{Obsah}
\VytvorObsah
\cleardoublepage

\pagestyle{plain}     
\setcounter{page}{1}  
\pagenumbering{arabic}


%

\chapter{Introduction}
\pagestyle{plain}     
\setcounter{page}{1}  
\pagenumbering{arabic}

As technology goes forward new and more powerful tool appears in each field of science. In case of nowadays astronomy and astrophysics in order to perform our work more effectively and efficiently, we are pushed towards cooperation with the computer science. Therefore, to run more powerful pipeline, the advantage of always improving programming languages need to be included.\

Old projects for reduction of astronomical data have been replaced with new ones using more precise programming codes. It is the case of Canada-France-Hawaii Telescope with ESPaDOnS echelle spectrograph. Upena pipeline, which used closed-source Libre-Espirt software, will be soon replaced by OPERA, Open source Pipeline for ESPaDOnS Reduction and Analysis. Disadvantage of closed-source Libre-Espirt is the fact, that it is available only at certain computer machines on Mauna Kea, Hawaii. Whereas, the use of OPERA pipeline can be applied to any other spectrographs in the world. \

Fortunately, my supervisor Petr Škoda from the Astronomical Institute in Ondřejov, got in touch with current developer of OPERA and started cooperation in order to apply it to Ondřejov Echelle Spectrograph attached to 2m Perek telescope in Czech Republic, OES. Although Ondřejov already has reduction tool for echelle data called Image Reduction and Analysis Facility, IRAF, it is not working sufficiently. Echelle data from OES has lines tilted with respect to the echelle orders, which is the problem not solvable for IRAF. That is reason why this thesis is concentrated on producing well functioning pipeline with OPERA software for OES data and analyzing the differences between data reduced by IRAF and OPERA. Final comparison of spectra from both tools will be plotted, described and analyzed in order to find the most suitable reduction software for Ondřejov Echelle Spectrograph.

\chapter{A~Brief Introduction to Spectroscopy}
\label{cha.intro}
Historically, it has been proved that for getting information about the celestial bodies so far away in the universe, it is useful to investigate incoming electromagnetic waves radiated from them. That is the field of study of spectroscopy. \
Generally, spectroscopy is a~scientific technique, which studies the interaction between electromagnetic radiation and matter. It is a~measurement of the radiation intensity as a~function of wavelength.\

Very famous experiment, made by Isaac Newton in 1666, offered explanation for this theory using the dispersion of sunlight on a~prism. He splitted white light to further undistributed colored parts with different index of refraction, that formed a~spectrum. \

By Thomas Young's inteferometry experiment was explained color assignment to specific wavelength. At the beginning of the 19th century a~talented glass maker Joseph von Fraunhofer found several dark lines in a~continuous spectrum of the Sun and with the use of telescope he was able to observe spectrum of the Moon, Mars, Venus and several other stars.\ 

The mystery of the dark lines was revealed 30 years later by Gustav Kirchhoff and Robert Bunsen and formed to Kirchhoff's laws of emission and absorption. It is used to determine the composition of celestial bodies from the spectral lines, characteristic for each element.\
Kirchhoff's law describes equivalent ratio of emissive and absorptive power of rays with the same wavelength at thermal equilibrium.
\begin{equation}
\frac{\epsilon_{\lambda}(T)}{\kappa_{\lambda}(T)} = \mathrm{constant},
\end{equation}
where the coefficient of emission is $\epsilon_{\lambda}$, the coefficient of absorption is $\kappa_{\lambda}$, the temperature $\mathrm{T}$ and the wavelength is $\mathrm{\lambda}$. Additionally, the Fraunhoffer dark lines and stellar lines can be explained by the Kirchhoff's laws:
\begin{enumerate}[label=(\alph*)]
\item the hot,dense solid objects emit the light with a~continuous spectrum  
\item hot, diffuse gas emits the light only at the specific waveleghts and forms emission spectrum
\item and finally, a~cool gas absorbing just specific waveleghts of obscured source shows near-continuous spectrum with dark emission lines.
\end{enumerate}



The simplest early spectroscope was a~prism. However, the light beam passing through the prism is scattered and absorbed, in any case it can be ''lost". Lord Rayleigh demonstrated that diffraction grating represents preferable  tool for obtaining higher resolutions. Anyway, several problems raise up in this case, too. The difficulty to produce grating at such quality and the low efficiency as the light disperse into several orders.\

Later, the improvement was offered by Fraunhofer. His idea of gratings with many finely ruled apertures on the glass was modified by Lewis Rutherford, who ruled a~small number of gratings in metal. At the end of the 19th century the effective manufacturing at John Hopkins University produced ''blazed" gratings with ability of light diffraction up to 50~$\%$ into first order.\

After the interchangeable prisms and gratings in spectrograph of Lick refractor, the Cassegrain spectrograph was constructed. It enhanced several advanced techniques for flexed control and was used at the Cassegrain focus of the telescope. The light coming into the telescope was reflected by primary and secondary mirror to the Cassegrain focus and spectrograph. Then it passed through the slit and  the parabolic mirror collimated it to the diffraction grating. The spectrum was formed and focused by lenses onto the photographic plates (nowadays, it is mostly used a~charged couple device, CCD).\

Furthermore, flexure problems were eliminated by construction of a~coudé spectrograph located at the coudé focus. The light was reflected by mirrors to the telescope axis,where the coudé slit room was located. Then the light beam passed through the slit of the spectrograph to the diffraction element forming spectrum onto the photographic plate.\

First, the prisms were used as a~diffraction element, later were replaced by blazed gratings and finally by Schmidt camera. In order to gain the highest resolution, the large gratings\footnote{With no limit for the space of the Cassegrain focus.}and the large photographic plates with the long focal length cameras were installed. Ideas of this methods were firstly used at the Mt Wilson spectrograph and after that it was spread to many observatories around the world and used ever since.


\chapter{The Echelle Gratings}
Generally, the most important factor of a~diffraction gratings is resolution. It is proportional to the total number of grating rules \textit{N} and the order of diffraction \textit{m}.\
When the highest resolution wants to be achieved, the main goal should be followed:
\begin{itemize}
\item increasing the number of grating rules
\item increasing the order of diffraction.
\end{itemize} 
Subsequently, the larger and more finely ruled grating had been used. As it was mentioned in previous section, the most effective setup was with the coudé spectrograph.\

However, blazed gratings achieve to disperse the light only into second and third order. Albert Michelson's experiments of an ''échelon" consisted of a~small number of parallel plates made of glass. The resolving power of the order of one million was achieved in 1933 by Williams. At that time conventional grating ''échelette" ruled on metal was designed by R.W.Wood.\

And finally, nowadays well known and all around the world used ''echelle" grating was described by Harrison in 1949. The word ''échelle", in French, means stairs or ladder.  There are several differences between conventional and echelle grating. Mainly, echelle represents an compact and high-resolution instrument. This statement is result of few properties it holds. Echelle has less grooves per millimeter than échelette and is used at high angles into high diffraction orders, so that it is more efficient and has low polarization effects over the large spectral intervals.\

The product of such high-resolution grating represents overlapping orders separated by crossed low dispersion element. Usually, prism, grism or exceptionally another grating is used as a~cross-disperser.
\section{Grating equation}
This subsection is offering important echelle grating equations. Firstly, it is concentrated on a~standard diffraction grating. Fig. \ref{fig.schema} is showing its schema in both pictures. Picture \textbf{b)} showing the standard grating with an incident angle $\gamma$ with respect to the facet normal.

 \begin{figure}[H]
      \centering
      \begin{minipage}{0.45\linewidth}        
              \includegraphics[width=0.9\linewidth]{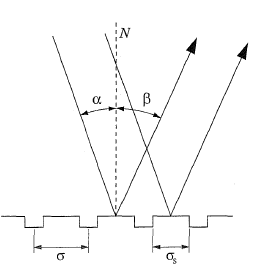}
              \caption*{a)}
			   \label{fig.schema}
       \end{minipage}
      \hspace{0.05\linewidth}
      \begin{minipage}{0.45\linewidth}
              \includegraphics[width=0.9\linewidth]{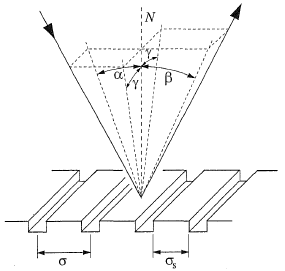}
              \caption*{b)}
			   \label{blaze}
      \end{minipage}
      \caption{Schematic diagrams of a~standard diffraction grating: \textbf{a)} all beams parallel to grooves  and \textbf{b)} incident angle $\gamma$ with respect to $N$.\cite{ech}}
     \label{fig.schema}
   \end{figure}

Grating is composed of reflective ''tooth" or grooves with a~normal $N$ and parallel rules with a~spacing \textit{$\sigma$}. The light incidents at the angle $\alpha$  and reflects at the angle $\beta$. Total path difference between groves forms a~standard grating equation: 
\begin{eqnarray}
m\lambda = \sigma \left(sin\alpha \pm sin\beta \right)
\label{eq.grading}
\end{eqnarray}
and for the inclination angle $\gamma$ additional term will appear:
\begin{equation}
m\lambda = \sigma \left(sin\alpha \pm sin\beta \right)~cos\gamma.
\end{equation}

This condition posses the plus sign, if the diffracted beam is on the same side of the grating as the incident beam. Otherwise, it is negative.\

Huygen's principle stands that each groove facet with width of $\sigma\mathrm{_s}$ is a~source of se-\
condary spherical wavelets.\

Constructive interference will occur, if diffracted beam path differs by an integer wavelength number $(m= \pm 1, \pm 2, \dots)$. It will interfere destructively if $m$ is $\pm 1/2, \pm 3/2,\ldots$. \

As the most of the light is reflected into a~zeroth order, wavelengths start to overlap. The solution is offered by blazed reflection gratings.
The position of the grating can be tilted so, that the incident ($\alpha$) and diffracted ($\bar{\beta}$) beam are at almost the same angle, in order to direct them towards preferable direction. For instance, towards the specific diffraction order.
\begin{figure}[h!]
\centering
\includegraphics[scale=5,width=280pt]{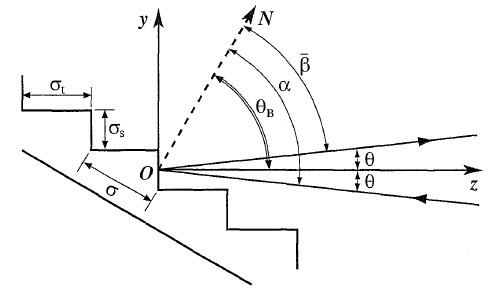}
\caption{Profile image of grating and blaze angle $\theta_B$.\cite{ech}}
\label{blaze}
\end{figure}

Profile of the tilted grating is shown in Fig. \ref{blaze}. The two angles, incident $\alpha$ and diffracted $\bar{\beta}$, are measured from grating normal and third angle $\theta_\mathrm{B}$ is measured from facet normal to the light beam. Next relations can be applied:
\begin{eqnarray}
\alpha &= \theta_\mathrm{B} + \theta \nonumber \\
\bar{\beta} &= \theta_\mathrm{B} - \theta \\  
\Rightarrow \alpha + \bar{\beta}&= 2~\theta_\mathrm{B}.
\label{eq.blaze}
\end{eqnarray}
Taking under account very polished surfaces, the peak intensity in diffraction matches the specular reflection of the groove facet, $\alpha$ and $\bar{\beta}$ are equal. Only single outgoing angle is produced and it's called blaze angle $\theta_\mathrm{B}$.\

 Moreover, the blaze wavelength $\lambda_\mathrm{B}$ can be defined as :

\begin{eqnarray}
m\lambda_\mathrm{B} &=& \sigma \left(sin\alpha \pm sin\bar{\beta} \right) cos \gamma \nonumber \\
 &=& \sigma ~ sin\theta_\mathrm{B} ~ cos\theta ~ cos \gamma
\end{eqnarray}
Efficiency of grating operation is limited
to only viable modes, where $\alpha > \bar{\beta}$ or $\alpha \approx \bar{\beta}$. Let's consider the Littrow condition, when $\alpha = \bar{\beta}$ that means $\theta = 0$ and quasi-Littrow, when $\gamma\neq 0$. Considering this condition, the optical depth of a grating $\sigma_\mathrm{t}$ is:\\ 
\begin{equation}
\sigma_\mathrm{t} = \sigma sin\theta_\mathrm{B}
\end{equation}
and facet width is given by:
\begin{equation}
\sigma_\mathrm{s} = \sigma cos \theta_\mathrm{B}
\end{equation}
That was last step of determining the order of interference for diffracted light:
\begin{equation}
m = \frac{2\sigma_\mathrm{t}}{\lambda}
\end{equation}

Grating with large blaze angle are nothing less than echelle grating. The name ''R-number" is given by the tangent of the blaze angle. For instance, R2 grating with an angle $\theta_B = 63.4^\circ$.\

\section{Cross-disperser}
\begin{figure}[h!]
\centering
\includegraphics[scale=5, width=200pt]{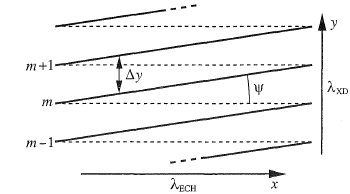}

\caption{The schema of cross-dispersion and cross-disperser grism.\cite{ech}}
\label{fig.cross}
\end{figure}

As mentioned before, echelle grating produces overlapping orders (m$\gg$~1), which is separated by the second dispersive element, prism, grims or grating (Fig. \ref{fig.cross}). The two Figs. (\ref{prism}
 and \ref{grading}) underneath give a~comparison between orders imagined by prism and grating. Grism represents the combination of prism and grating in order to get light at chosen wavelength passed straight through.

 \begin{minipage}{0.9\textwidth}
      \centering
      \begin{figure}[H]
      \begin{minipage}{0.45\textwidth}
          \begin{figure}[H]
          	\centering
              \includegraphics[width=150pt]{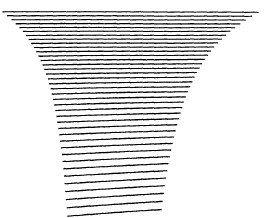}
              \caption{a)}
              \label{prism}
          \end{figure}
      \end{minipage}
      \begin{minipage}{0.45\textwidth}
          \begin{figure}[H]
          	\centering
              \includegraphics[scale=5,width=150pt]{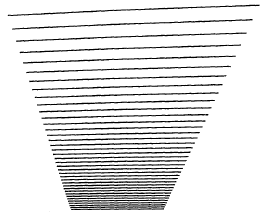}
              \caption{b)}
              \label{grading}
          \end{figure}    
   \end{minipage}
\caption*{Orders with: \textbf{a)} prism and \textbf{b)} grating.\cite{ech}}
\end{figure}
  \end{minipage}

\subsection*{Order separation}
Additionally, the condition of the separation between orders can be applied:
\begin{equation}
\Delta y = f_{\mathrm{cam}} \frac{\mathrm{d}\beta}{\mathrm{d}\lambda_{\mathrm{XD}} \Delta\lambda_{\mathrm{FSR}}},
\end{equation}
where $f_\mathrm{cam}$ is the focal length of spectrograph camera, $\frac{\mathrm{d}\beta}{\mathrm{d}\lambda_{\mathrm{XD}}}$ is the angular dispersion of the cross-disperser and $\Delta\lambda_{\mathrm{FSR}}$ is the free spectral range.\\

The free spectral range is an changing wavelength from an order \textit{m} to the next $(m \pm 1)$:\noindent
\begin{equation}
\Delta\lambda_{\mathrm{FSR}} =  \frac{\lambda}{m},
\end{equation}
Each wavelength in order \textit{m} is also present in orders $(m \pm 1)$.\ \noindent

Let's consider blaze wavelength $\lambda_\mathrm{B}$ in equation for free spectral range:\noindent
\begin{equation}
\Delta\lambda_{\mathrm{FSR}} = \frac{\lambda_{\mathrm{B}}^2}{2\sigma ~sin\theta_\mathrm{B}~cos\theta~cos \gamma},
\end{equation}
and apply it to the order of separation:\noindent
\begin{equation}
\Delta y = f_{\mathrm{cam}} \frac{d\beta}{d\lambda_{\mathrm{XD}}} \frac{\lambda_{\mathrm{B}}^2}{2\sigma ~sin\theta_\mathrm{B}~cos\theta~cos \gamma}
\end{equation}

\section{Resolving power}
The resolving power is very important property of any spectrograph. The smallest di\-ffe\-rence in wavelengths, $\Delta\lambda$, between two spectral lines, that are able to be distinguished at wavelength $\lambda$, defined as:
\begin{eqnarray}
R = \frac{\lambda}{\Delta\lambda},
\label{eq.res}
\end{eqnarray}
where $\lambda \approx \lambda_1 \approx \lambda_2$. 

\subsection{Angular dispersion}
The angular dispersion applied into the relation, will give: \noindent
\begin{equation}
R = \frac{\lambda}{\delta \beta} \frac{\mathrm{d}\beta}{\mathrm{d}\lambda},
\end{equation}
where $\delta\beta$ is an angular width between two wavelength in the dispersed beam, followed by angular dispersion $\left(\frac{\mathrm{d}\beta}{\mathrm{d}\lambda}\right)$.

\begin{figure}[h!]
\centering
\includegraphics[scale=5,width=200pt]{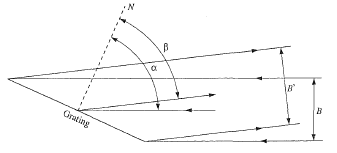}
\caption{Anamorphic magnification on dispersed light from echelle grating.\cite{ech}}
\label{fig.anamorphic}
\end{figure}

Importantly, the effect of anamorphic magnification~(Fig. \ref{fig.anamorphic}) needs to be considered. The source, viewed from grating at angular distance $\delta\alpha$, has after dispersion angular separation $\delta\beta$:\noindent
\begin{equation}
\delta\beta = \delta\alpha\frac{\mathrm{d}\beta}{\mathrm{d}\alpha}
\end{equation} 
An anamorphic magnification is described in following equation:\\ \noindent
\begin{equation}
r = \bigg\vert \frac{\mathrm{d}\beta}{\mathrm{d}\alpha} \bigg\vert = \frac{cos\alpha}{cos\beta}
\end{equation}

Furthermore, amortification can be applied to the relation of resolving power and it becomes:\\ \noindent
\begin{equation}
R = \frac{\lambda}{\delta \alpha} \frac{\mathrm{d}\alpha}{\mathrm{d}\lambda} 
\end{equation}

Or, when the relation\noindent 
$\frac{{\delta \alpha}}{\mathrm{d}\lambda} = \frac{sin\alpha+sin\beta}{\lambda~cos\alpha}$ is applied:\\ \noindent
\begin{eqnarray}
R &=& \frac{1}{\delta\alpha}~\frac{sin\alpha + sin\beta}{\lambda~cos\alpha} \\
&=& \frac{1}{\delta\alpha}~\frac{2tan\theta_B}{1-tan\theta_B~tan\theta}
\end{eqnarray}

In addition, the diffraction limitation on resolving power needs to be included too. Grating of length \textit{L}, with the number of grooves \textit{N} 
across it, gives the limit:\noindent
\begin{eqnarray}
m\lambda &=& \frac{L}{N} \left(sin\alpha + sin\beta \right)~\mathrm{and} \\
R &=& mN.
\end{eqnarray}
In diffraction limit, when the collimated beam size is \textit{B} equal to ($L~cos\alpha$) and the limit of angular size of a~slit is $\approx$ $\lambda/B$.\ \noindent

\subsection{Linear Dispersion}

Linear dispersion determines the range of wavelengths, that will fit onto the detector at the focal plane of the spectrograph.
Let's consider $\alpha$ to be the same for all wavelengths. Consequentially, for the given $m$ the differentiation of the grating equation \ref{eq.grading} will give:
\begin{eqnarray}
m = a~cos\beta~ \frac{\mathrm{d}\beta}{\mathrm{d}\lambda} \Rightarrow \frac{\mathrm{d}\beta}{\mathrm{d}\lambda} = \frac{m}{a~cos\beta},
\end{eqnarray}
where \textit{a} is the number of lines per mm.\
Than the spectrum is linearly dispersed at x-direction in camera's focal plane \textit{f$_{\mathrm{cam}}$}

\begin{eqnarray}
\frac{\mathrm{d}x}{\mathrm{d}\lambda}= f_{\mathrm{cam}}~ \frac{\mathrm{d}\beta}{\mathrm{d}\lambda} =\frac{m~f_\mathrm{cam}}{a~cos\beta}
\end{eqnarray} 

The resolving power can be derived from reciprocal form of the linear dispersion in \textup{\AA ~}/mm and with well known size of a~pixel on CCD. It will give the dispersion per pixel, the smallest distinguishable wavelength $\Delta\lambda$ (in \textup{\AA}) in equation for resolving power (\ref{eq.res}). \

Two previous sections were made from \cite{ech} and \cite{ostlie}.\\ \noindent

\chapter{The Echelle Spectrographs}
\label{eche}
The aim of this chapter is to introduce and compare echelle spectrographs all around the world. The closer look is taken at the Ondřejov Echelle Spectrograph (OES) at 2m telescope in Ondřejov, Heidelberg Extended Range Optical Spectrograph and the Echelle SpectroPolarimetric Device for the Observation of Stars at Canada-France-Hawaii Telescope on Hawaii. The first spectrograph, as the main topic of this thesis. The second one, as an inspiration for constructing OES  and the last as the main reason why the new project for reduction and analysis was produced.\\

\section{Ondřejov Echelle Spectrograph}
\label{cha.oes}
In 2000, the research team of stellar astronomers from the Astronomical Institute of the Czech Academy of Sciences in Ondřejov under the lead of P. Koubsky decided to construct the Ondřejov Echelle Spectrograph (OES). The spectral region limitation was the main reason why a~standard coudé spectrograph was not sufficient enough. As an inspiration served echelle spectrograph HEROS (Heidelberg Extended Range Optical Spectrograph) from
Germany. \

HEROS is a~fiber-fed spectrograph designed and build by  Andreas Kaufer, et al. in 1994. The light from the telescope is carried by 10m fiber, which has an 100 micron core diameter. Since the end of fiber represented the point source, slit is not required.
Echelle grating has 31.6 grooves/mm with an blaze angle 63.4 deg~(R2 grating) and contains two separate camera channels with different wavelength ranges, red~(365-565nm) and blue~(580-835nm). That means HEROS is able to cover at once around 400nm-long part of a~spectrum, almost 10 times more than can cover standard spectrograph in Ondřejov, with circa twice bigger resolving power.\

In testing process, HEROS was attached to the Cassegrain focus and more than 1500 spectra were obtained, although, some problems had arisen. Mainly, by merging the orders reflected to variability of wave-like structures of continuum. The problem remained after flat-fielding and was not overcome.\\

After 4 years of gathering experiences was Ondřejov echelle spectrograph fully functional. Although, HEROS was great inspiration some changes in construction of OES needed to be done. \

OES is a~slit spectrograph. For the fiber-fed mode the fiber would need to be 20m long, as the light travels from primary or Cassegrain focus.\

\begin{figure}[h!]
\centering
\includegraphics[scale=4,width=150pt]{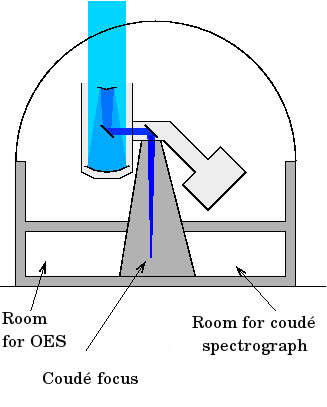}
\caption{Location of Ondřejov Echelle Spectrograph.\cite{soes}}
\label{fig.loca}
\end{figure}

In fully operational mode, OES is stored in the room build symmetrically to the coudé focus of 2m telescope and to the room for standard spectrograph\footnote{The difference is that standard spectrograph has only one disperse unit (prism or grating)}~as it is illustrated in Fig.\ref{fig.loca}. Since in the coudé focus is situated a~plane mirror, it is possible to switch between two spectrographs within a~night.\

Additionally in this setup fiber would represent only redundant optical element.\

\subsection*{Schema of OES}

\begin{figure}[h!]
    \centering
   	\includegraphics[scale=5,width=200pt]{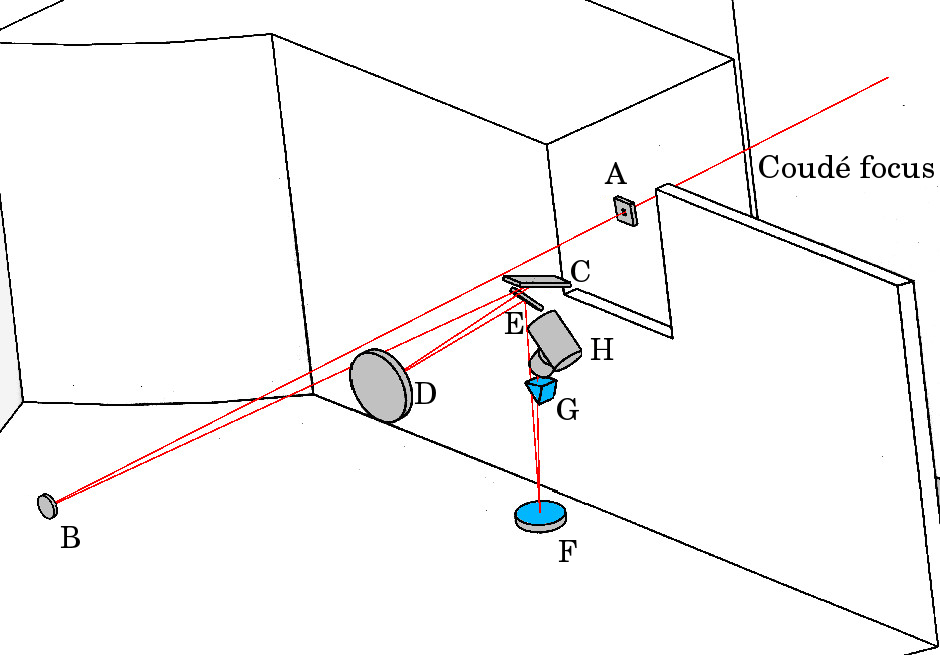}
\includegraphics[scale=5,width=200pt]{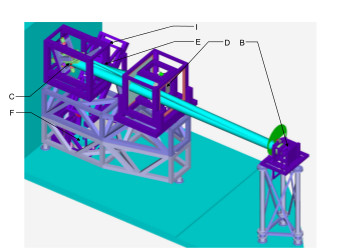}
\caption{
Schematic setup of OES: A) Slit, B) Collimator No.1, C) Echelle grating, D) Parabolic mirror, E) Additional flat mirror, F) Collimator No.2, G) Prism, H) Lens of camera, I) CCD detector (right picture).\cite{oes}}
\label{fig.test}
\end{figure}

The schema of the setup of important optical parts of spectrograph and the light beam passing through shows Fig.\ref{fig.test}. Additional information will be considered.\

Left hand side part of Fig. \ref{fig.test} shows setup for OES. The light beam, painted in red, is coming from the coudé focus through a~0.6\textit{mm} slit~(A), at distance 500\textit{mm} from the front edge of the grating (C). Next, the beam is converted to a~parallel beam by the collimator No.1 with the focal length $\approx$ 4600\textit{mm}. The optical diameter of the collimator and as well the diameter of produced beam is approximately 150\textit{mm}. The most important optical part, echelle grating (C), was constructed by Milton Roy, has a~blaze angle of 69$^\circ$ (tan$\approx$2.6), 54.5 grooves/mm and the size 154$\times$408mm. Produced spectrum, with overlapping orders, is pictured on a~parabolic mirror (D) with focal length  approximately 1100mm and it is reflected on small mirror(E) with an 45$^\circ$ angle of inclination and next to the second collimator~(F), with almost 600 mm smaller than the focal length of parabolic mirror (D). Last optical element represents cross-dispeser (G), prism from a~glass Schott LF5\footnote{LF means ''light flint"} with an vertex angle 54.5$^\circ$. Spectral orders, divided by prism, are filling up the 30$\times$10.5mm area in focus of 200mm camera(H)-Canon EF. The beam incidents at the angle 46.8$^\circ$. The mean deviation at 480nm is 38.6$^\circ$.\cite{oes} \

\begin{figure}[h!]
\centering
\includegraphics[scale=5,width=200pt]{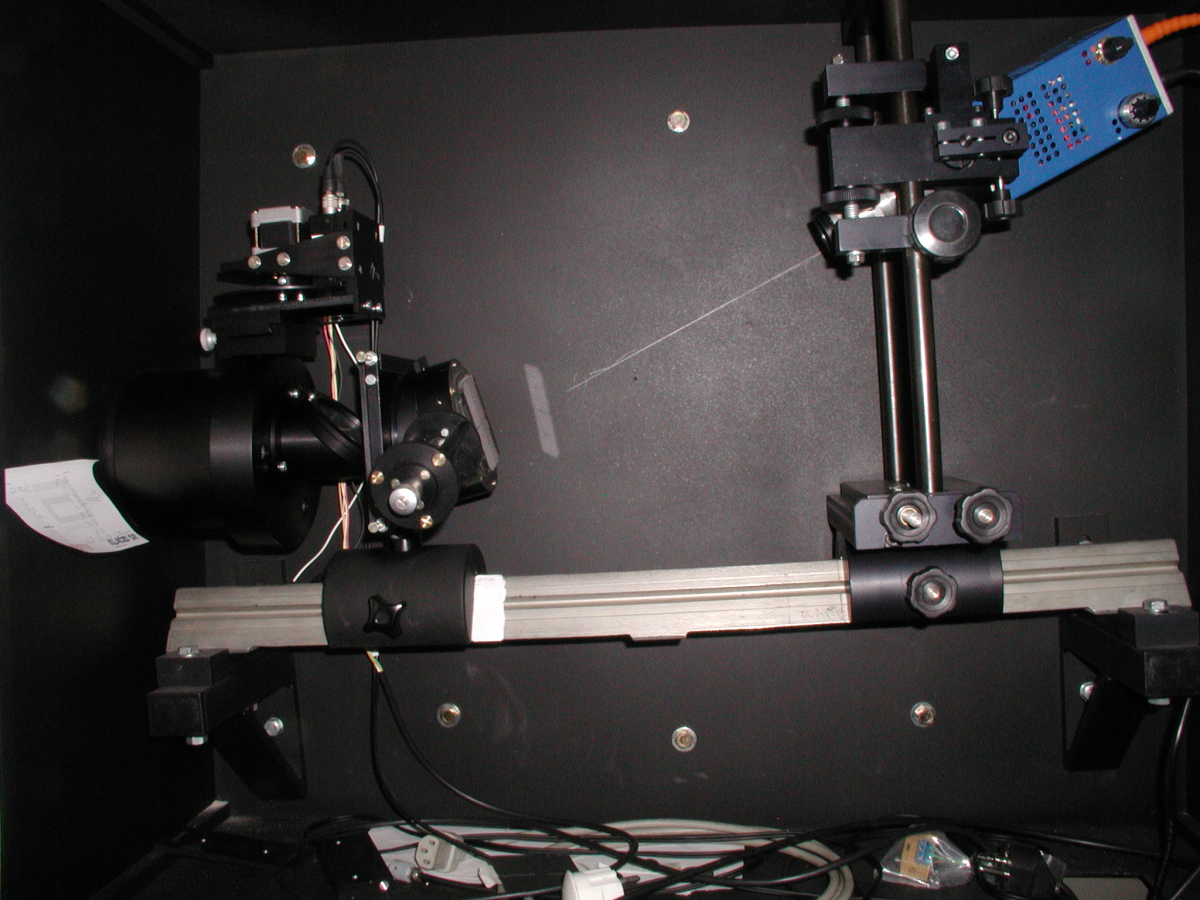}
\includegraphics[scale=5,width=200pt]{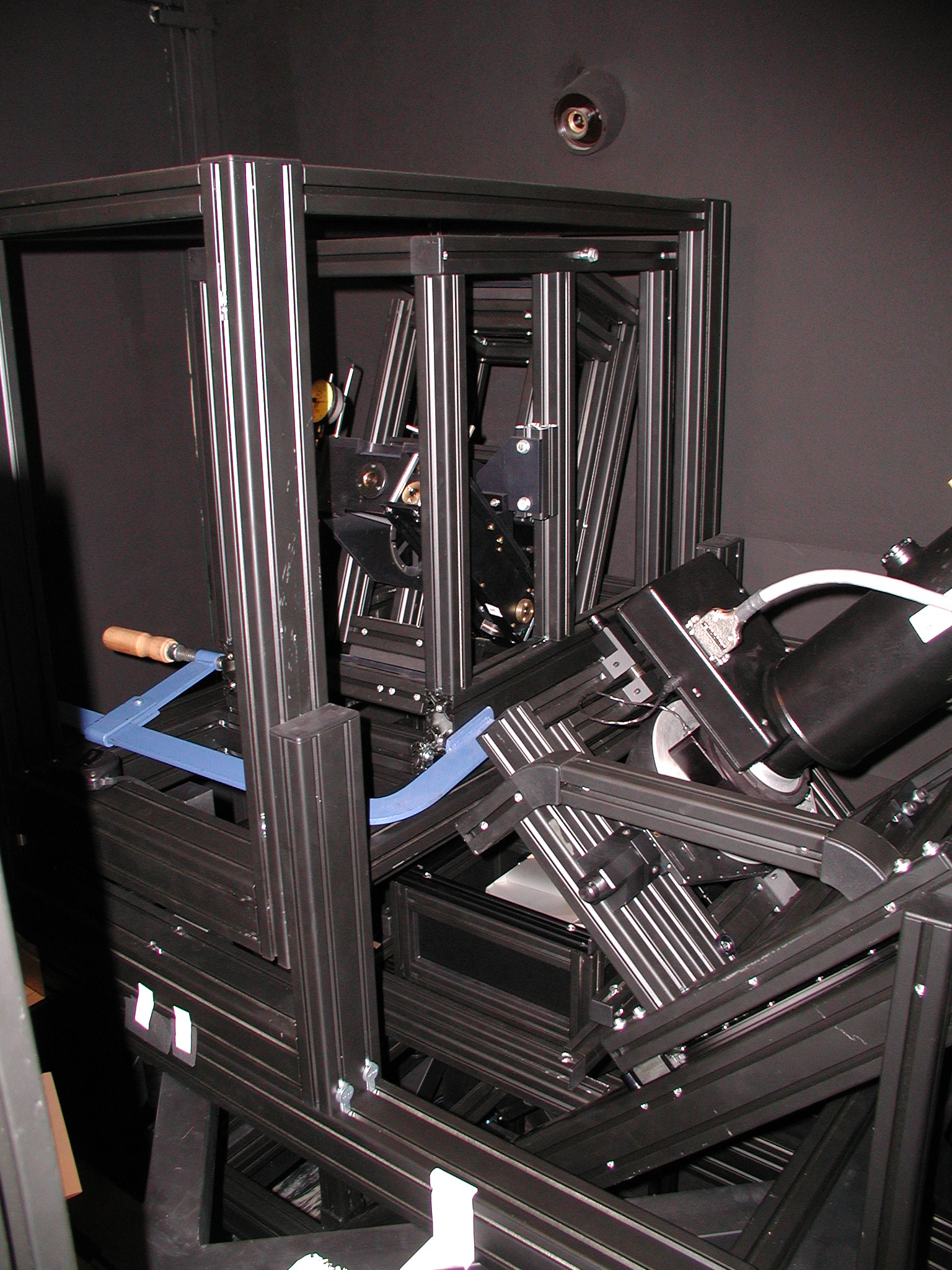}
\caption{Ondřejov Echelle Spectrograph. Pictures were obtained from server of Stellar Department.}
\label{fig.oes_p}
\end{figure}

\subsection*{Iodine Cell} 
\label{iodine}
\begin{figure}[h!]
\centering
\includegraphics[scale=5,width=300pt]{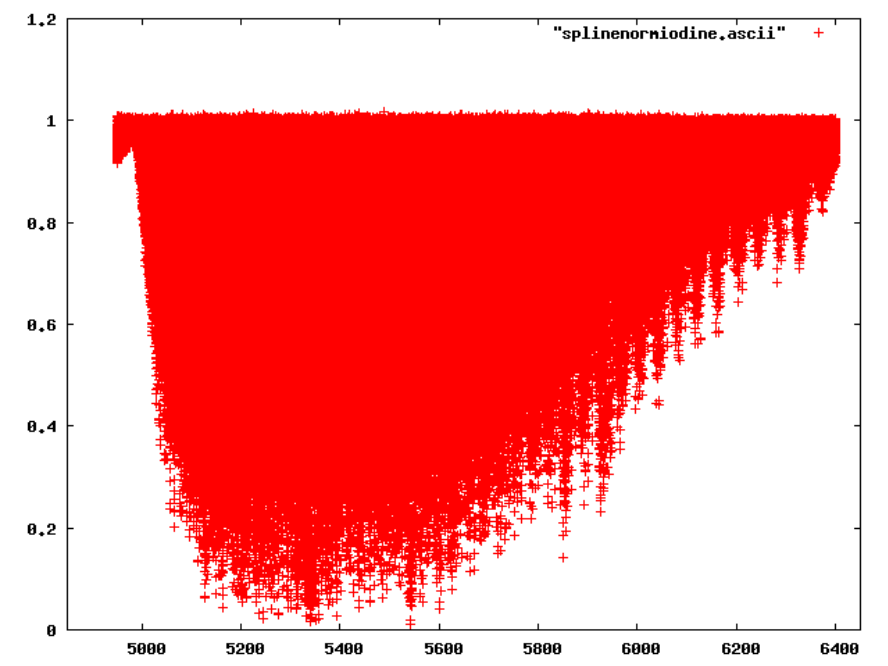}
\caption{Iodine cell spectral range. \cite{iodin}}
\label{fig.io}
\end{figure}

OES can use an Iodine cell for high-precision spectroscopy as well. The Iodine (I$_2$) absorption cell is a vacuum sealed glass jar with Iodine crystals and is wrapped with heat ribbon to warm it on working temperature. Iodine sublimes at a temperature 35$^\circ$C, while the working temperature is typically at 60-80$^\circ$C, and produces a spectrum of deep and very narrow absorption lines in range of the waveleghts between 4800~\AA ~ and 6100~\AA ~, whereas the highest absorption is in range from 5100~\AA ~till 5600~\AA ~as demonstrates Fig. \ref{fig.io}. This generated spectrum represents very stable zero-velocity reference system. \

Although, some observations with Iodine cell in Ondřejov have been already performed, further research and adjustments are required in this field. Sample data with Iodine cell are presented in the chapter \ref{cha.comp}, which includes final extracted spectra from both OPERA and IRAF.   
\subsection*{Spectral range}
The Ondřejov echelle covers visible spectral range from 3870~\AA ~ to 10100~\AA \footnote{Although infra-red part is influenced by very bad fringing and Thorium-Argon calibrations in this part are missing or are not applicable.} in 42 orders recorded on 2048$\times$2048 EEV chip in an Dewar jar. The distance between orders is 12 pixels in red part of a~spectrum and 27 pixels in blue. Additionally, one pixel has size 13.5 $\mu m$.\\
\subsection*{Spectral resolving power}
Determination of resolving power is a complex problem. It can be precisely determined only by measurement with laser. However, for approximate estimation is sufficient enough to just calculate practical value from two pixel resolution or theoretical value from parameters of instrument, particularly grating and detector.\

In practice, it is possible for OES to get spectral resolving power for each line separately. An example reduction of V1679 Cygni by M. Šlechta shows, that the wavelength 5000\AA ~ ~belongs to 21st order. The value of CDELT1~$ =0.0407419$ is found in the header of V1679 Cyg. Center is around 5010~\AA ~, where the value 0.041~\AA ~is the distance between centers of two neighbouring pixels and those two pixels are 0.082~\AA ~apart. Spectral resolving power for two pixels is:
\begin{equation}
R = \frac{\lambda}{\Delta\lambda} = \frac{5010}{0.082} = 61100
\end{equation}
Considering H$\alpha$ line, the order 38 is investigated. Center of this order is located around 6590~\AA ~ ~with the distance between two pixels 0.05362. What gives the resolving power for H$\alpha$:
\begin{equation}
R = \frac{\lambda}{\Delta\lambda} = \frac{6590}{0.011} = 61450
\end{equation} 
\\

Theoretical resolving power was derived by Pavel Koubsky in the article \cite{oes}. Considering estimated linear dispersion from observation (values of CRVAL in header) on the value 2,4 \textup{\AA}/mm at 5000\textup{\AA}, the width of the slit is 600 $\mu m$, than image would have 40 $\mu m$ (almost more than three 13.5 $\mu m$ pixels) in the camera, what leads to the resolution 51 600 at 5000 \textup{\AA}. Two pixel resolution is 77000.\
 Important is to note, that those are just approximate values. After extraction of stars by OPERA, its statistics showed the value not higher than 56000. \\

Now we focus on the spectrograph, which was the inspiration for the production of the new reduction tool.

\section{ESPaDOnS}
\label{cha.esp}
ESPaDOnS is an \textbf{E}chelle \textbf{S}pectro\textbf{P}olarimetric \textbf{D}evice for the \textbf{O}bservation of \textbf{S}tars at 3.6m Canada-France-Hawaii Telescope. Design and the construction was made by a~team of 15 scientists, engineers, technicians, and administrators at he Observatoire Midi-Pyrénées in France inspired by FEROS spectrograph.\

Two units forms ESPaDOnS, bench-mounted fiber-fed echelle spectrograph and on telescope attached Cassegrain unit, which includes polarimeter, calibration unit and camera, all connected with 30m of fiber. Spectroscopic analysis is concerned with tree modes: standard spectroscopy, spectropolarimetry and for the very faint objects standard spectroscopy with subtracted spectrum of the sky.
Beside the uniqueness in measuring the magnetic topogies of stars, ESPaDOnS is used for galactic objects, stellar and planetary studies.\\

Spectrograph is stored in thermal enclosure on third floor of Coudé room. This way the affects on spectrograph by temperature and pressure fluctuation are minimized.
\begin{figure}[h!]
\centering
\includegraphics[scale=5,width=200pt]{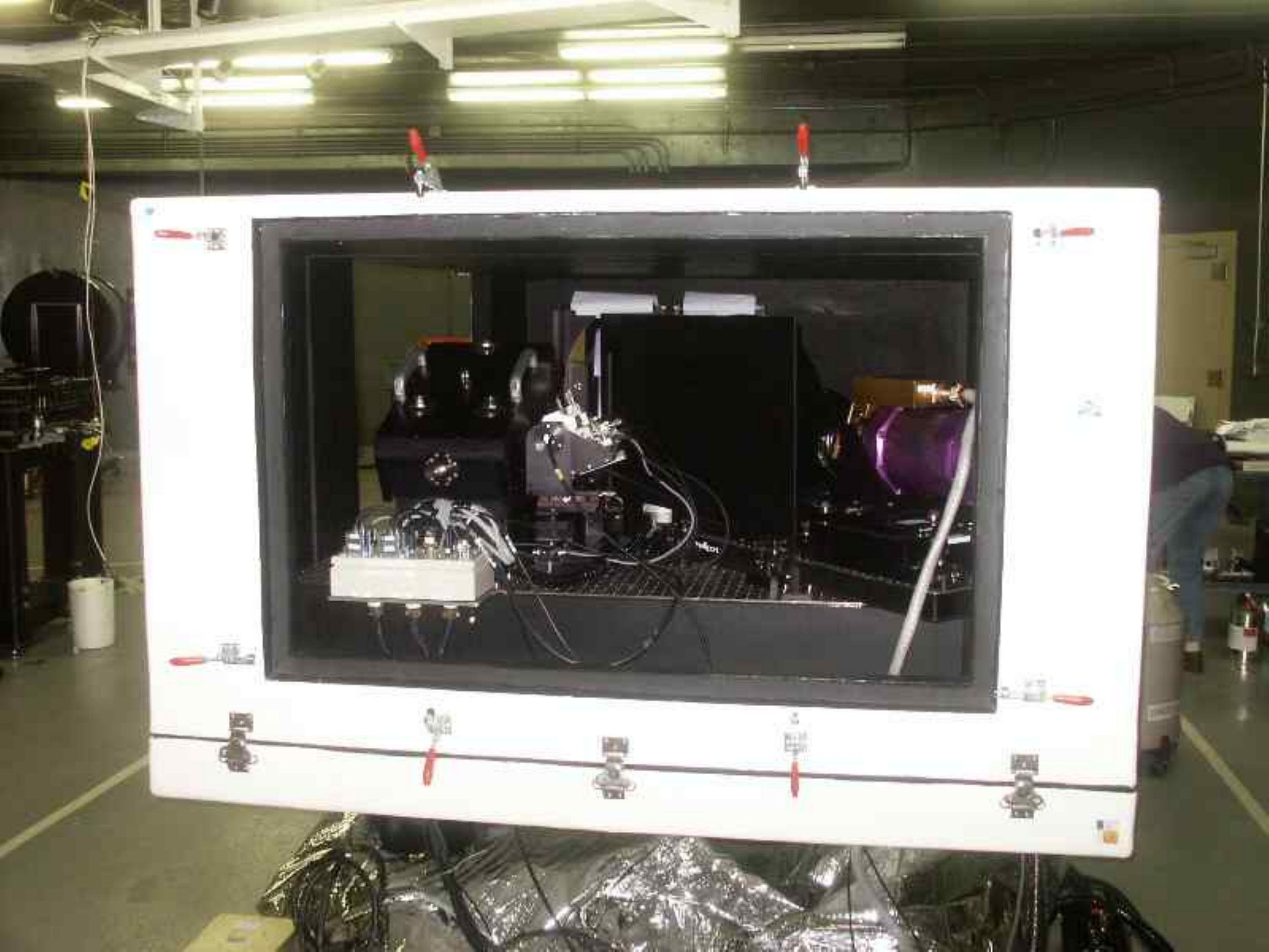}
\includegraphics[scale=5,width=200pt]{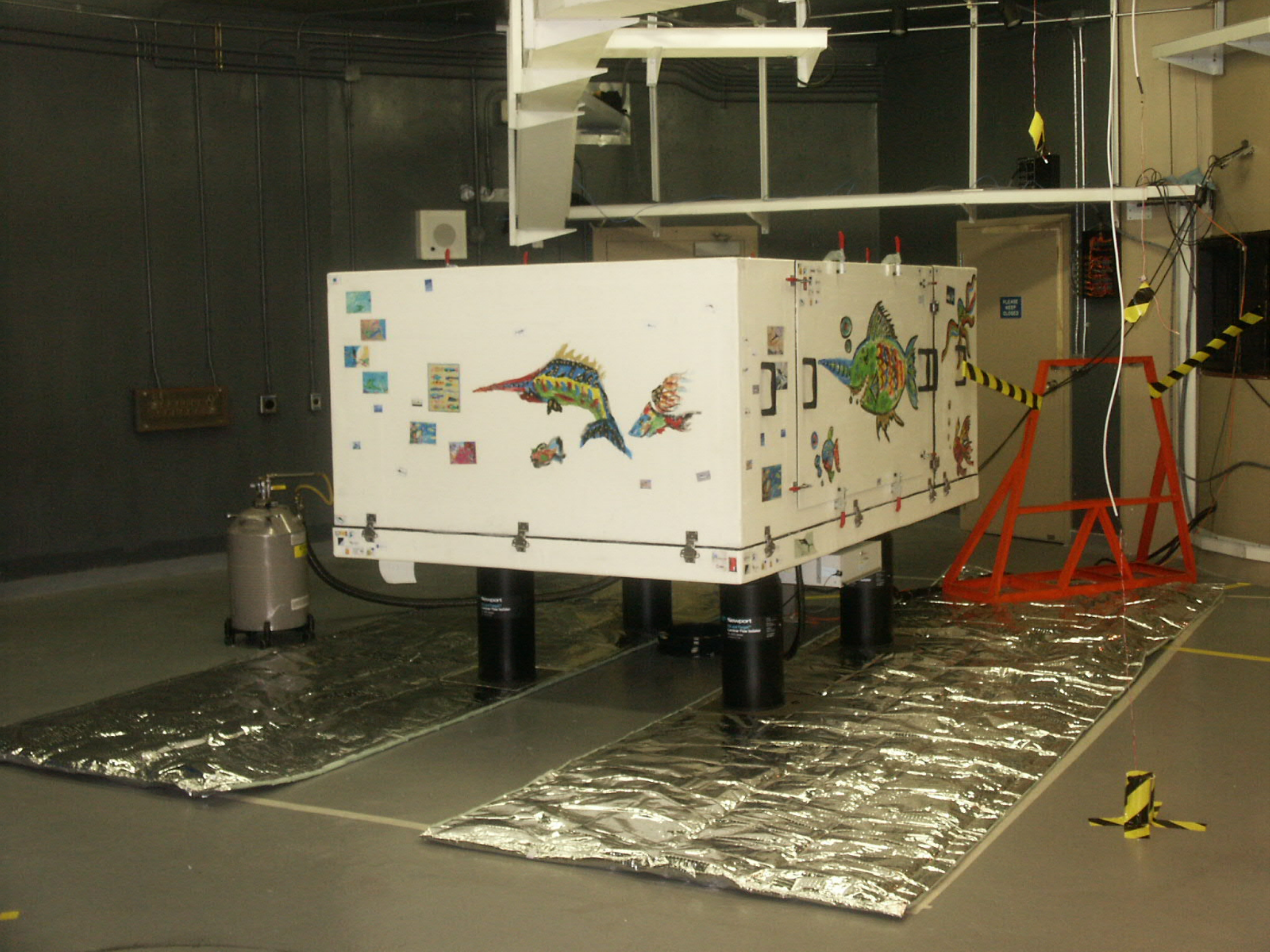}
\caption{ESPaDOnS in its thermal enclosure, in the 3rd floor Inner Coudé room.\cite{espadons}}
\label{fig.tran}
\end{figure}

Interesting fact is, that with the longest 270m fiber ever made is ESPaDOnS connected to its 8m neighbour telescope. This project is called GRACES, Gemini Remote Access to CFHT ESPaDOnS Spectrograph, and works in the same modes, except polarimetry. Access is granted when spectrograph is not used by the CFHT. In 2016 the bilateral agreement will be executed. The CFHT community will be able to use their telescopes, Gemini North or South, with ratio 3/20 nights.  \cite{newe}

\subsection*{Schema of ESPaDOnS}

\begin{figure}[h!]
\centering
\includegraphics[scale=5,width=250pt]{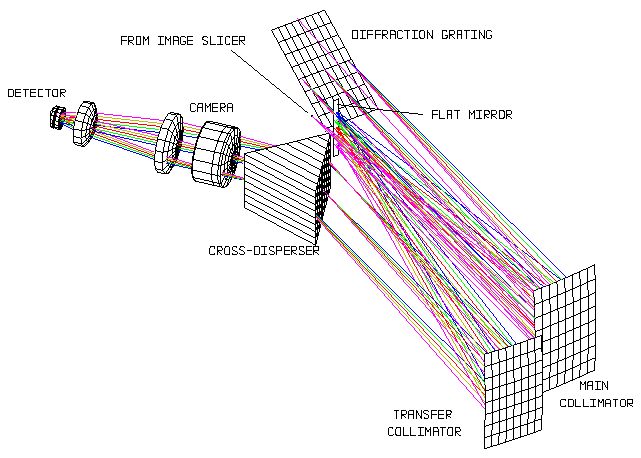}
\caption{The schema of spectrograph.\cite{esp_conf}}
\label{fig.sesp}
\end{figure}

This echelle spectrograph is working in dual pupil and quasi Littrow configuration. Setup of optical parts is shown on Fig. \ref{fig.sesp}.\

The light is fed by optical fibers to the Bowen-Walraven image slicer made of three individual prisms for transforming the circular images into the narrow strip at the spectrograph slit level. The statistics claims, that about 40 $\%$ to 45 $\%$ of stellar photons made it from the telescope to the spectrograph. After slicing, the light is collimated by the main collimator to the diffraction grating and back to it, forming the first overlapping spectrum near to the flat mirror. It's role is to send the beam to the transfer collimator and to the cross-disperser, in this case the prism. Thereafter the final image is recorded onto 2048x4096 15 micron pixel CCD detector of the fully dioptric camera with 388mm focal length and with free diameter of 210mm.\

\subsection*{Spectral range and resolution}
ESPaDOnS covers not only the visible part of the spectrum from 370nm to 740nm, but also continues to near infrared part till 1050nm\footnote{Three small gaps in infrared are included.} in a~single exposure. It provides spectral coverage of 40 orders from the order 61 to the order 22. Echelle R2 grating has 79 grooves per mm, the size 200x400mm.    \

Spectral resolution differs depending on mode that is used. Considering sky and standard spectroscopy and spectropolarimetry the resolution reaches 68 000, whereas for standard spectroscopy only the resolution increases, up to 81 000.


%
%
%

\section{Comparison of Echelle spectrographs}
The short overview of the echelle spectrographs is presented in the Table \ref{tab:hsk}. It includes samples from all around the world. Two German spectrographs from Heidelberg and Tautenburg, SOPHIE in Haute-Provence Observatory in France, Netherlands UES on one of the Canary Island, La Palma, attached to the William Herschel Telescope and Belgium HERMES at the Mercator Telescope and at last ESO facility HARPS at 3.6m and FEROS at 2.2m telescope in La Silla, Chille.\

HERMES and SOPHIE have  two modes available with different spectral resolving powers. Firstly, HERMES has high resolution mode~(HRF) with $R\sim$ 85 000 for high efficiency along with simultaneous wavelength reference mode~(LRF) with \textit{R} up to 62 000 for high instrument stability. In the case of SOPHIE, high resolution mode~(HR) with $R\sim$ 75 000 achieved by 40$\mu$m slit and HE as high efficiency mode concentrating on fainter objects with 25 000 smaller spectral resolution.\

Cross-dispersing element is indispensable for each echelle spectrum. Table \ref{tab:hsk} shows, that in most cases this role represents prism. Remaining spectrographs are using grism, so called grating prism. \

ESO's HARPS holds first place for the highest resolving power. On the other hand, German HEROS takes the last place. Whereas the longest spectral range is covered by UES. Certainly, the range is not completely continuous, small gaps in between two parts of the spectrum are included.\

As a matter of fact, HARPS takes the highest places especially in field of low-mass exoplanetary research, as its name implies. Statistics from 2010 stands, that HARPS discovered 23 of 28 exoplanets with mass 18 times smaller than Earth mass. The rate is still increasing. Unquestionably, because of sub-ms$^{-1}$ precision of radial velocity determination. On the other hand, new-generation spectrograph ESPRESSO is optimized for discoveries of non-active rocky planets of few Earth masses.
\begin{table}[h!]
\begin{small}
\centering
\caption{The comparison of echelle spectrographs.\cite{hermes},\cite{sophie},\cite{harps},\cite{heros},\cite{soes},\cite{espadons},\cite{tau}.}
\label{tab:hsk}

\begin{tabular}{p{1.95cm}p{3.7cm}p{2.1cm}p{1.9cm}p{1.8cm}p{1.1cm}}

&  &  Resolving & Spectral & Cross-&Fibre- \\
Short cut& Name&power &range[\AA]&disperser&fed\\

\hline\hline
OES&Ondřejov Echelle S.\ &50000& 3870-10100&prism&no\\
\hline
ESPaDOnS&Echelle SpectroPolarimetric Device for the Observation of Stars &81000&3700-10500&prism&yes\\ 
\hline
HEROS& Heidelberg Extended &20000 & 3450-8650& prism&yes\\
&Range Optical S.&&&&\\
\hline
FEROS& The Fiber-fed Extended &48000 & 3500-9200& prism&yes\\
&Range Optical S.&&&&\\
\hline
HARPS&High Accuracy RV & 115000& 3780-6910 & grism &yes\\
&Planet Searcher&&& grism&yes\\
\hline
ESPRESSO&Echelle S. for Rocky Exoplanets and Stable &HR:134000  MR:59000 &3800-7800&prism&yes\\
& Spectroscopic Observations&UHR:200000&&&\\
\hline
UES &Utrech Echelle S. & 54000 &3000-11000& prism&no\\
\hline
& Echelle Tautenburg &67000 &3400-9270& grism&no\\
\hline
HERMES& High Efficiency and Resolution Mercator &HRF:85000 LRF:62000 & 3770-9000&prism &yes\\
&&Echelle S.&&&\\
\hline
SOPHIE & Spectrograph for Observation of Phenomena of stellar interiors and&  HE:75000 HR:40000& 3870-6940&prism &yes\\
&  Exoplanet&  & & & \\
\hline

\end{tabular}
\end{small} 

\end{table}

\newpage

The data gathered in the Table (\ref{tab:hsk}) suggests that optical fiber for transmitting the light from the telescope to the spectrograph are frequent component of those instruments.\

\begin{figure}[h!]
\centering
\includegraphics[scale=4,width=150pt]{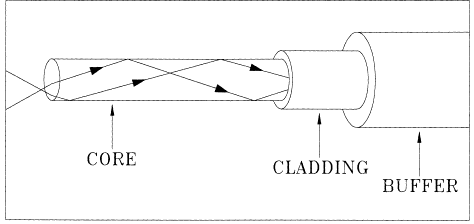}
\caption{The construction of optical fiber. \cite{fiber}}
\label{fig.fiber}
\end{figure}

Since 1970 the optical fiber are often used in spectroscopy. Optical fibers (\ref{fig.fiber}) are made of glass or plastic with higher reflective index than cladding, which is covering the fiber. Once the light is reflected at larger angle than the critical angle, it is totally internally reflected. This principal is essential for functioning of fibers. The buffer is for mechanical protection.\

To main advantages of this instrument belongs high flexibility, more mechanically stable spectra, multi-fibers observation spectroscopy, integral field spectroscopy and sightly higher spectral resolution than slit, because of circular apertures. On the other hand, some problems by using fiber have arisen. Internal transmission, absorption and circular aperture losses, focal-ratio degradation\footnote{The emergent light has lower focal-ratio than input light is more spread out in an angle. }, variance of light flux along the observation, constant core diameter not adjustable to seeing condition. For telescope with different than coudé or Nasmyth feed, are optical fibers very convenient.\\

%
%
%

One of the most important capabilities of echelle spectrograph, high resolution, offers possibility to explore important fields of stellar physics unreachable for standard spectroscopes. \

Gradually, larger and larger telescopes are constructed, what extends the capability of spectrograph to reach to the  fainter objects. As the photon collective power increases too, the high precision spectroscopy is in demand. With precise spectroscopy the instrumental effects on spectra can be reduced, what results in providing highly repeatable observations over long timescales. Most importantly, the determination of the position and shapes of spectral lines are included, mainly the high precision measurements of radial velocity. ESO-ESA\footnote{ESO stands for European Organization for Astronomical Research in the Southern Hemisphere and ESA for European Space Agency} working group's report from 2005 states, that those velocity instrumentation are for further following-up of the astrometric and transit detection methods of exoplanets, to ensure detection by this second method and determining its true mass. Along the support experiments are provide to improve radial velocity mass detection limits.

The search for terrestrial planets in habitable zones is considered to be one of the leading topics of the next decades in Astronomy. Moreover, it represents important application of the purpose of high dispersion spectrographs. That is the reason why next chapter of this thesis will be dedicated to the topic of exoplanets.

\chapter{Exoplanets}
\label{cha.exo}
\section{Definition and discoveries}
Exoplanet (extra solar planet) is known as a~planet, that orbits other star than Sun.\ 

That brings out a~question, how is actually the word ''planet" defined?
As it was agreed in February 2003 by the Working Group on Extrasolar Planets (WGESP) of the International Astronomical Union, ''planet" by its definition is an object with true masses below the limiting mass for thermonuclear fusion of deuterium.\footnote{Calculated for objects of solar metallicity to be $\sim$ 13 M$_\mathrm{J}$ (Jupiter masses)} On the other hand the substellar objects above this limit are defined as ''brown dwarfs", not considering their location of the origin or the process of the formation. \ 

Existence of exoplanets was predicted in 18th century by Immanuel Kant. In those times it was just an idea, because there was no possible way how to prove it. 
First exoplanet was discovered by Alex Wolszczan and Dale Frail in 1992 unexpectedly around a~pulsar. Three years later Michel Mayor and Didier Queloz of the University of Geneva found an exoplanet around the star 51 Pegasi b \footnote{The naming convention for exoplanets stands of the star's name, which the exoplanet wandering about, followed by ''b",''c"...in the order of the planet's discovery.}, main-sequence star, that has triggered the boom of the exoplanets discoveries.\\

Fig. (\ref{fig.graph_ex}) below shows the numbers of discoveries from 1992 till nowadays with both direct and indirect methods, that can be used to find new extrasolar planets. As of the October 2015, about 1968 planets had been discovered in 1248 planetary systems, including 490 multiple systems. This data is available on The Extrasolar Planets Encyclopedia in their catalog. The biggest increase was observed around year 2014, mainly because of discoveries made by Kepler spacecraft. It will be described in the section about the transit detection.

\begin{figure}[h!]
\centering
\includegraphics[scale=5,width=350pt]{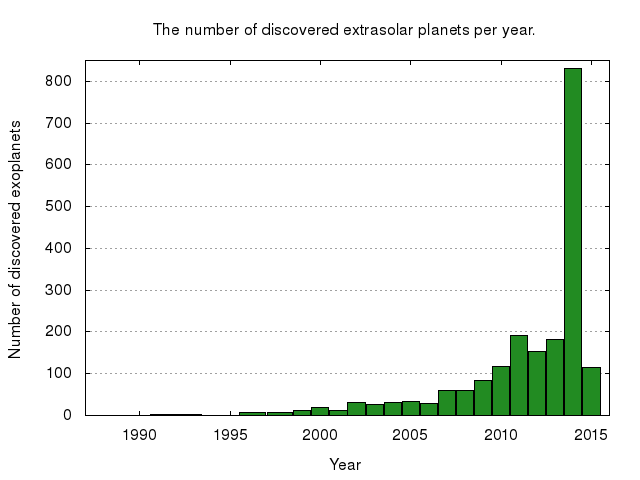}
\caption{The number of discoveries per year for all methods.}
\label{fig.graph_ex}
\end{figure}
\newpage
\section{Detection Techniques}

Exoplanets are really difficult to detect and imagine, as a~result of a~small angular separation and a~large brightness contrast between the planet and the star. The problem with contrast can be partly solved by looking at the dimmer stars such as white and brown dwarfs. Although, there are no discoveries whatsoever around white dwarfs, but there were found several planets around brown dwarfs by the Very Large Telescope.\
\subsection{Direct Method}
Usually, all of the exoplanets discovered by direct method are widely separated from the star and also have bigger mass then Jupiter. The majority of this planets have high temperatures, because they radiate in infrared wavelength so they are fainter in visible light. In some cases\footnote{The formation of young giant planets during accretion phase.} is preferable to make observation at 656,281~nm, well known wavelength for hydrogen line, H$\alpha$, because planets are more luminous at this wavelength than in near-infrared. That is the reason, why more H$\alpha$ surveys are undertaken \cite{ha} .\

Currently, several surveys exist for direct-imaging detection of exoplanets, namely Gemini Planet Imager, VLT-SPHERE and SCExAO.

On the other hand, there is a~possibility to detect the planet near the planet's Planck peak, what opens up the problem, that none of an achievable telescope would be sufficient enough for longer wavelengths, because of the diffraction limit of an angular resolution.\\

\subsection{Indirect Method}
\begin{figure}[h!]
\centering
\includegraphics[scale=5,width=350pt]{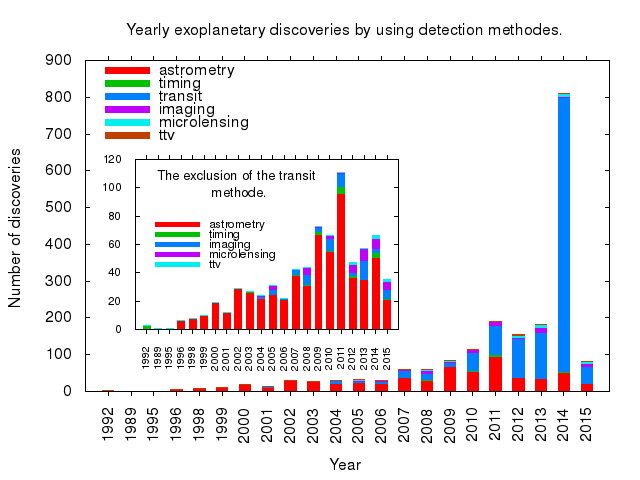}
\caption{Exoplanetary discoveries per year for different detection methods. Data from \cite{cex}.}
\label{fig.graph_meth}
\end{figure}

More successful approach is an indirect detection method. It takes advantage of the influence of the planet on dynamics or brightness of the host star.\
The most effective methods are:  
\begin{itemize}
\item the radial velocity or Doppler method
\item the astrometry
\item the planetary transits
\item the transit-timing variation (ttv)
\item the timing
\item the microlensing
\end{itemize}
Fig. (\ref{fig.graph_meth}), constructed by the data from \cite{cex} with gnuplot script (Appendix \ref{hist}), shows the comparison between exoplanetary discoveries by all known methods, in both direct and indirect way. The highest peak contains data from 2014 gained by transit method, as mentioned before, therefore additional graph is embedded with only astrometry, timing, imaging, microlensing and ttv.

\newpage
\subsubsection{The Radial Velocity Detection}
The RV method is mostly covered by high-precision measurement from echelle spectrograph, as mentioned before.
\noindent It uses the detection of the radial velocity \textit{v} of the star with mass M$_\mathrm{*}$ as the planet with mass M$_\mathrm{p}$ rotates around their centre-of-mass \textit{a} at the distance $2a~M_\mathrm{p}/\left(M_\mathrm{p} + M_\mathrm{*}\right)$ from the star. Let's assume that M$_\mathrm{*}\gg$~M$_\mathrm{p}$ and the angle between the normal of the plane of planet's orbit and the flow line between the star and the observer is the inclination \textit{i}, than the condition for the velocity:
\begin{equation}
v^2 = G \frac{\left(M_\mathrm{p}~sin~i~\right)^2}{2M_\mathrm{*}~a}
\end{equation}
The useful results of this method is that it is able to determine $(M_\mathrm{p}~sin~i)$ and planet's orbit eccentricity \textit{e} from the shape of time variation of velocity.\

On the other hand, important disadvantages rises up with unknown inclination angle of the orbit, so it is possible to determine only the lower limit on the planet's mass.\

It is mostly used for G and K~type, low mass stars, closely orbited by larger mass planets, in order to get the largest vision field of stellar velocity and it should offer the best possibility how to distinguish between the natural turbulent velocity in a~photosphere and studied radial velocity due to planetary orbit.
\subsubsection{Astrometry}

Astrometry is measuring the time changes of the position of the star in the sky. Planet's gravitational influence, causing reflex motion of the star, can be recognized.\
It is especially useful for describing and completing properties known from the other detection methods, for instance the absolute mass and orbital inclination of the planet.\

Let's consider the star and the planet moving around the same center-of-mass, than the reflex amplitude of the star gives:
\begin{equation}
a_\mathrm{*} = a_\mathrm{p}~\frac{M_\mathrm{p}}{M_\mathrm{*}},
\end{equation}
where $a_\mathrm{*}$ is the distance of the star from the centre-of-mass a~the same for the planet  ($a_\mathrm{p}$).\

This method is the most suitable for the systems near from the Earth, because it measures an actual angular position of parent star.\

The European Space Agency mission Gaia, launched at the end of 2013, contributes as well to astrometric detection. It's microarcsecond-precision of stellar positions will help to investigate the properties gained by another methods. It aims to detect over ten thousand Hot Jupiters\footnote{Hot Jupiters(or epistellar Jovians) are one type of exoplanets with Jupiter's characteristics. However, the very close orbit around the host star, causes much higher surface temperatures than for Jupiter. Bellerophon, or 51 Pegasi, is well known representative of this class.} and will affirm the existence of Jupiter-size planets from Doppler method \cite{gaia}.

\subsubsection{Transit Detection}
Gaia is used for transit detection only in conjunction with ground-based observation. Otherwise, it would rather give results for adventitious timing of the transit, because the time sampling of Gaia is weakly distributed \cite{gaia}.\

Transit method consists of measuring the decrease of the intensity of parent star as planet transits in front of its disk as seen in Fig. \ref{fig.tran}.
\begin{figure}[h!]
\centering
\includegraphics[scale=5,width=250pt]{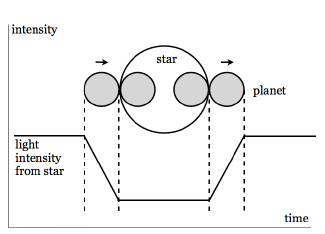}
\caption{Schematic illustration of the transit of planet in front of the star.\cite{e_a}}
\label{fig.tran}
\end{figure}

Some requirements for this method are raising up.
\begin{enumerate}
\item The instrument detecting the drop have to be sensitive enough to measure changes of luminosity. Less than 1~$\%$ for the larger planets, with the order of 10~$\%$ of the size of the star and even better sensitivity for smaller planets. \

For instance, 0.01~$\%$ for the Earth transiting the Sun for the alien observers.

\item The inclination angle \textit{i} of the planet's orbit should be very close to 90$^\circ$ for the ground-base observer. It is possible to determine \textit{i} and radius of the planet too, for the very large planets, that causing huge decrease of the luminosity of the parent star.\

The transit method defining the inclination angle can clear up the uncertainties in the mass determination, gained by radial velocity method.

\end{enumerate}

Kepler, the space mission launched by NASA, uses transits for discovering exoplanets. In 2014, the new statistical technique ''verification by multiplicity" was used. This method partly relies on the logical probability. Since for several stars is very difficult to orbit similarly to each other and stay in stable configuration, the drop of the light is rather caused by planets. Almost 92~$\%$ of all discoveries by all methods, that means 745 exoplanets was found by transit detection. 

\subsubsection*{Pulsar Timing}
A~pulsar, pulsating radio star, is a small ultra-dense highly magnetized rotating neutron star. The detection of planets around the pulsar are done by measuring the periodic time changes of the pulse.
The first detection of extrasolar planet was accomplished by Pulsar Timing method as was mentioned in section ''Definition and discoveries". \

On one hand, timing method can detect planets of the far smaller radii (than any other method can) and at much bigger distance from the pulsar. On the other hand, not many extrasolar planets have been found around pulsar, because the fraction of these objects is very low. Moreover, the high energy radiation of pulsars will not allow to any life form to be created.

\subsubsection{Transit-Timing Variation Detection}
Transit-timing variation method detect the variations in the timing of transits. Very sensible for observing multiple planetary system with mass of additional star similar to the Earth mass.\

\subsubsection{Microlensing}
Gravitational microlensing uses the fact that light from distant source is magnified by the gravitational field of the star (lens star) passing close to it or in the foreground. If lens star harbors planet, than it's own gravitational field contributes to lensing effect. 
This method is more useful for planets between the Earth and the center of the galaxy, because this part of Milky Way contains very large number of distant stars, that must to be continuously detected for observing the planetary microlensing contribution at an adequate rate.\

The advantage of microlensing over the other methods is the sensitivity to detect pla\-nets with larger orbits, namely around 1-10 AU away from the Sun. 
On the other hand, the process of microlensing is non-repetitive, because the justification will occur never again. Moreover, the planet is turned to be more distant, so no additional investigation of properties can be made.
%
Chapter \ref{cha.exo} is based on books \cite{e_a} and \cite{e_d}.


%
\chapter{Data Files from OES}
\label{cha.data}
The test data for this thesis is provided by Petr Škoda, a~regular staff member of Stellar Department at the Astronomical Institute in Ondřejov. It includes sample file (in FITS format) of the observed object, in this case Beta Lyrae traditionally named as Sheliak, the comparison spectrum (comp), the flat field (flat) and the zero :\\

c201307040002.fit $\Rightarrow$ object \ 

c201307040003.fit $\Rightarrow$ comp \

c201307040004.fit $\Rightarrow$ flat \

c201307040014.fit $\Rightarrow$ zero \\

\begin{figure}[h!]
\centering
\includegraphics[scale=5,width=200pt]{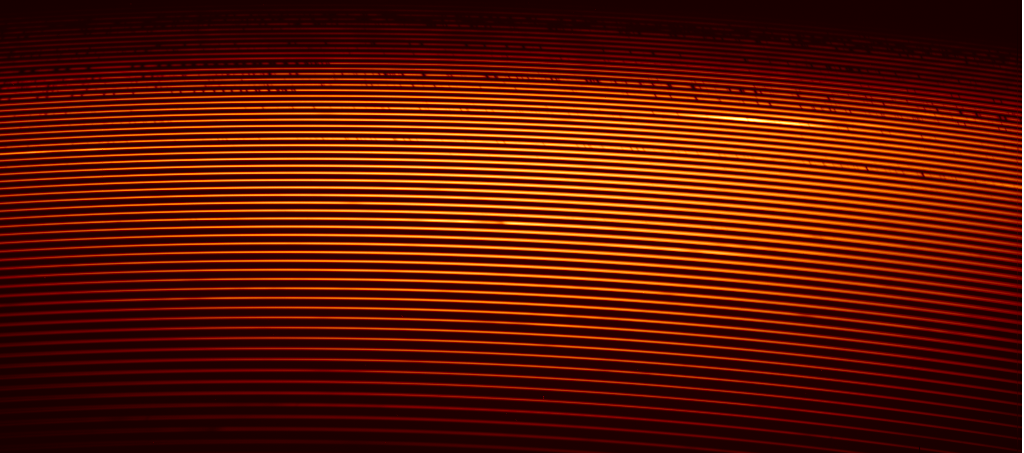}
\includegraphics[scale=5,width=200pt]{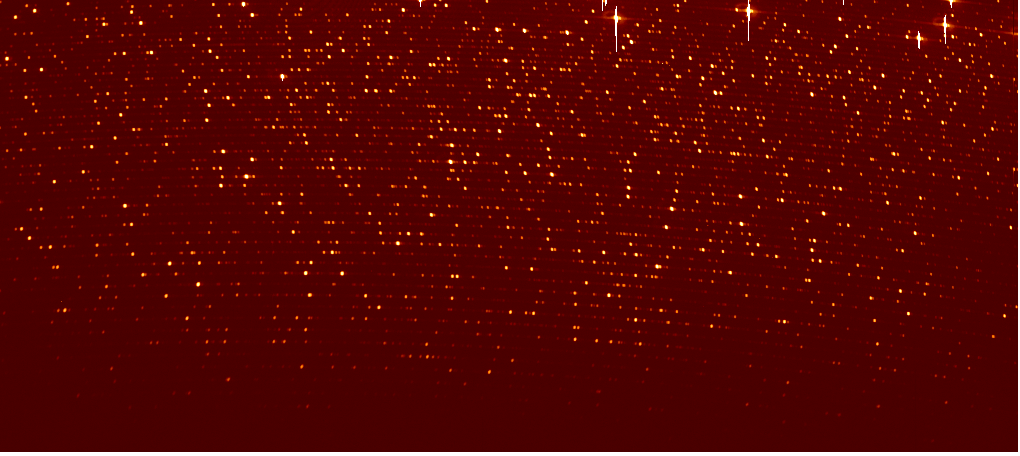}
\includegraphics[scale=5,width=200pt]{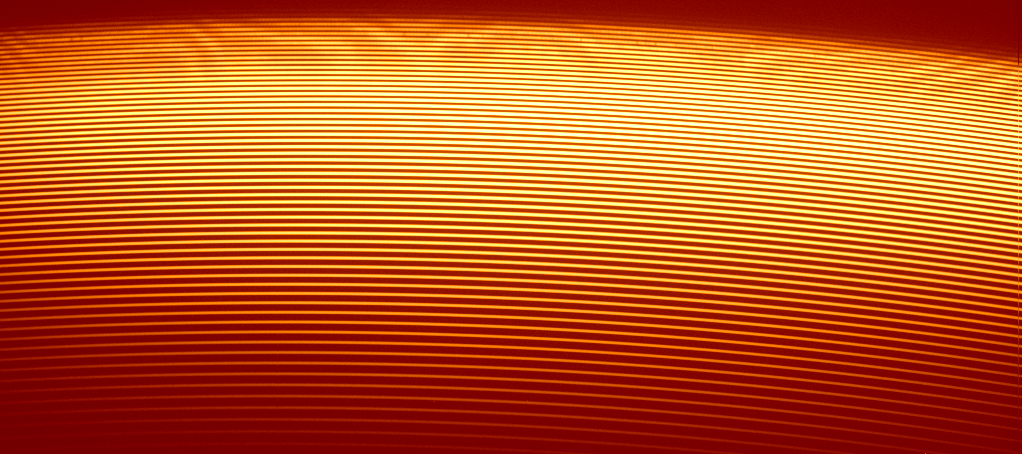}
\includegraphics[scale=5,width=200pt]{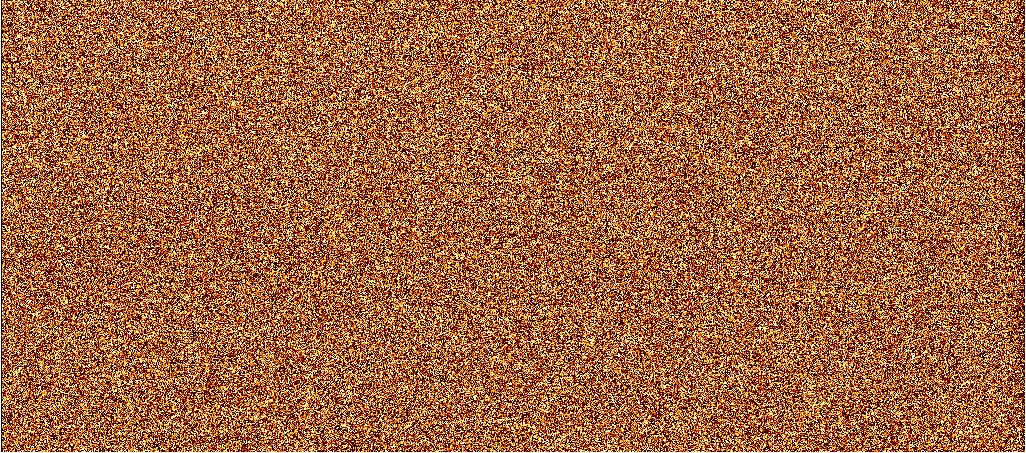}
\caption{Spectrum of star (top-left), comparison spectrum (top-right), flat field (bottom-left) and zero (bottom-right) displayed with ds9.}
\label{fig.ima}
\end{figure}
A~comparison spectrum is a~spectrum produced by a~lamp in laboratory conditions. Thorium-Argon is used in lamp in Ondřejov. Zero or bias belongs to the calibration images. When the spectrograph is closed, without any light, with zero exposure, the CCD camera chip is readout. Flat field images are produced by pointing telescope to the equally illuminated white space. The aim of this procedure is to recognize different sensitivities on the pixels on chip. The exposure time does not necessary need to be equal to the exposure of the star. Usually, from the flat fields are subtracted the zero images and the final spectrum is usually divided by flat field. But in case of echelle data this procedure is not possible. Zero is still subtracted, but easy division of the object is not possible in echelle setup. The reason includes the fact that the boundary of orders in cross-section profile of flat  are not aligned with those of stellar exposure so after division the strong singularities would appear. Because of this, each order needs to be flat-fielded separately only in 1D extracted spectrum.\

\begin{figure}[h!]
\centering
\includegraphics[scale=5,width=\textwidth]{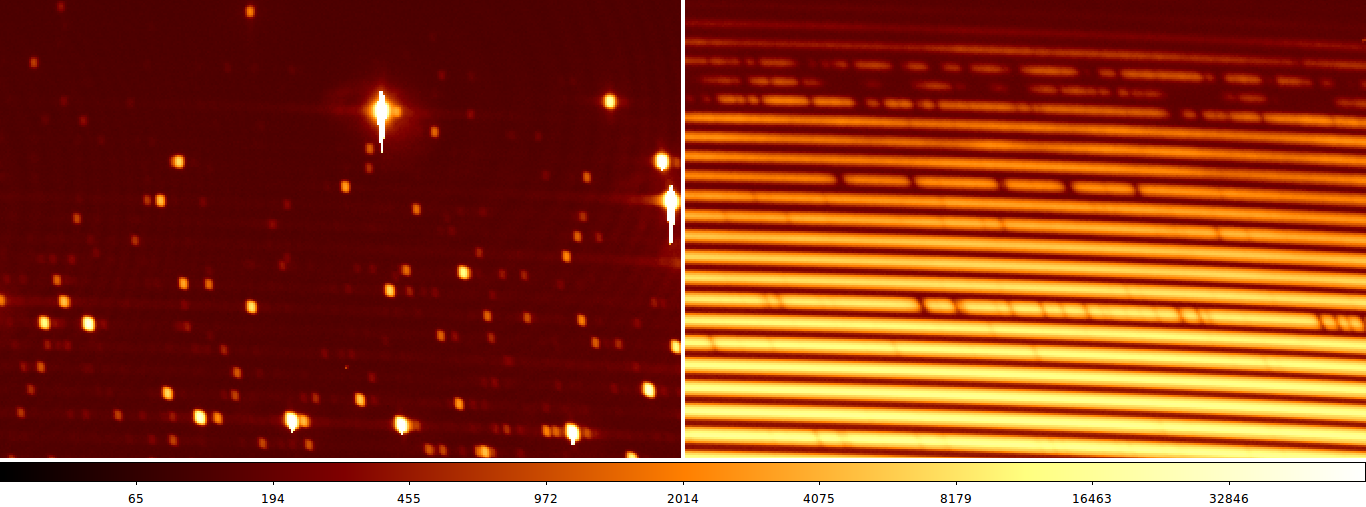}
\includegraphics[scale=5,width=\textwidth]{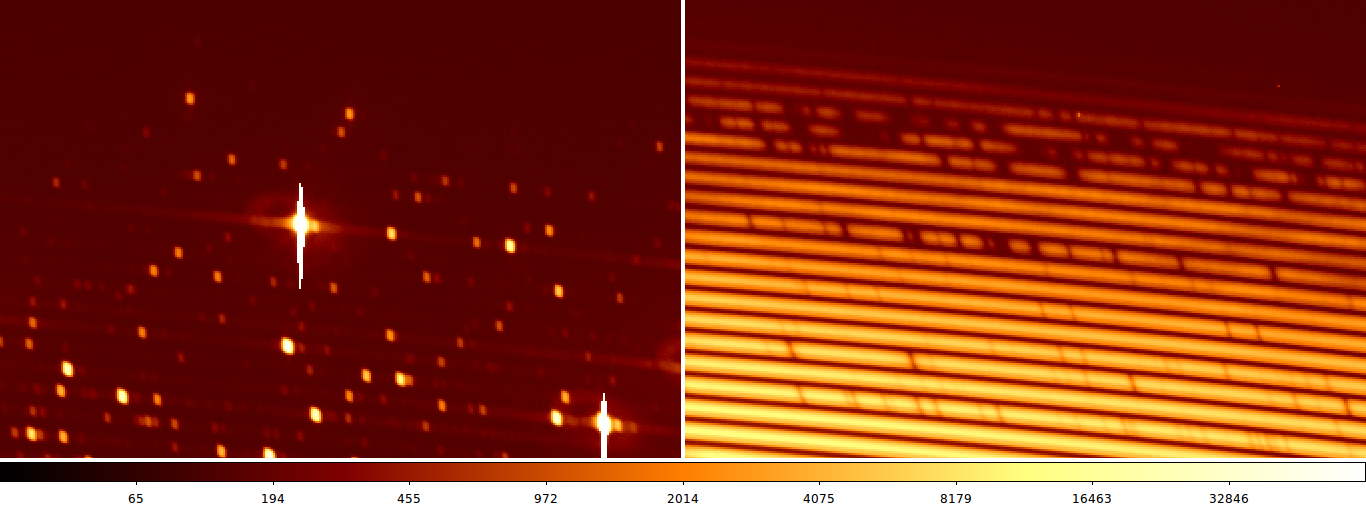}
\caption{Tilted lines in different part of the spectrum of star beta Lyrae (right) and straight lines of comparison spectrum (left).}
\label{fig.tilt}
\end{figure}

\begin{figure}[h!]
\centering
\includegraphics[scale=5,width=\textwidth]{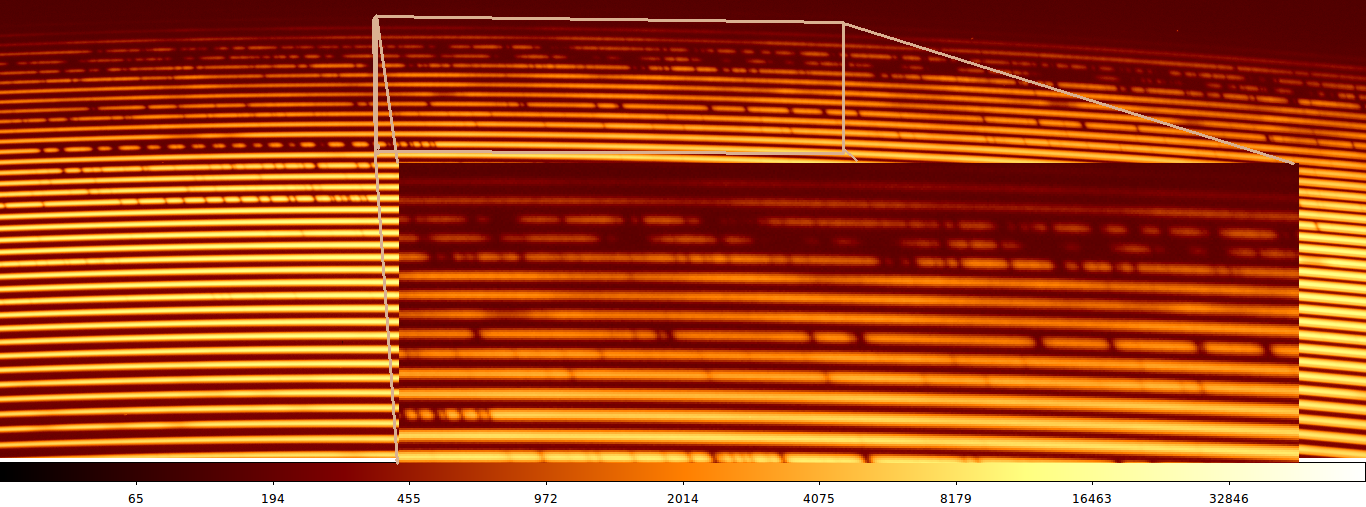}
\caption{Zoomed infra-red part of the spectrum of star beta Lyrae.}
\label{fig.tilt_gimp}
\end{figure}

Data were produced on 7th of April 2013 in Ondřejov. As it was mentioned in introduction, the OES spectrum shows the tilted spectral lines. This character is obtained in Fig. \ref{fig.tilt_gimp}. The comparison (Fig. \ref{fig.tilt}) of zoomed spectral lines in star spectrum and comparison spectrum is offered. The star spectrum is showing apparent tilt of line, whereas comparison spectrum in the same region only straight lines. This characteristic is explained by the deviation of the spectrograph from a perfect Littrow configuration, which causes spectral lines to tilt. Moreover, the angle varies along each order with position. \
It is not in IRAF capabilities to deal with those tilts and straighten them. New reduction tool is, therefore, demanded.

\begin{figure}[h!]
\centering
\includegraphics[scale=5,width=\textwidth]{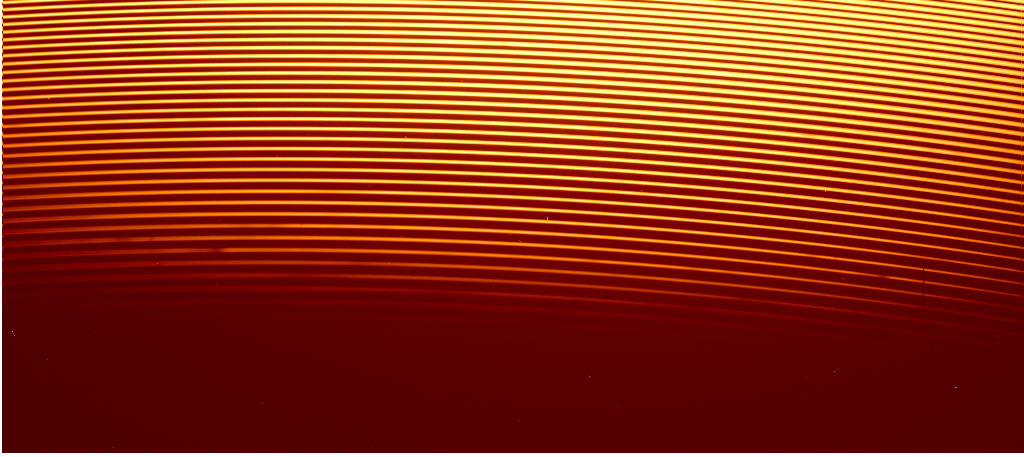}
\includegraphics[scale=5,width=\textwidth]{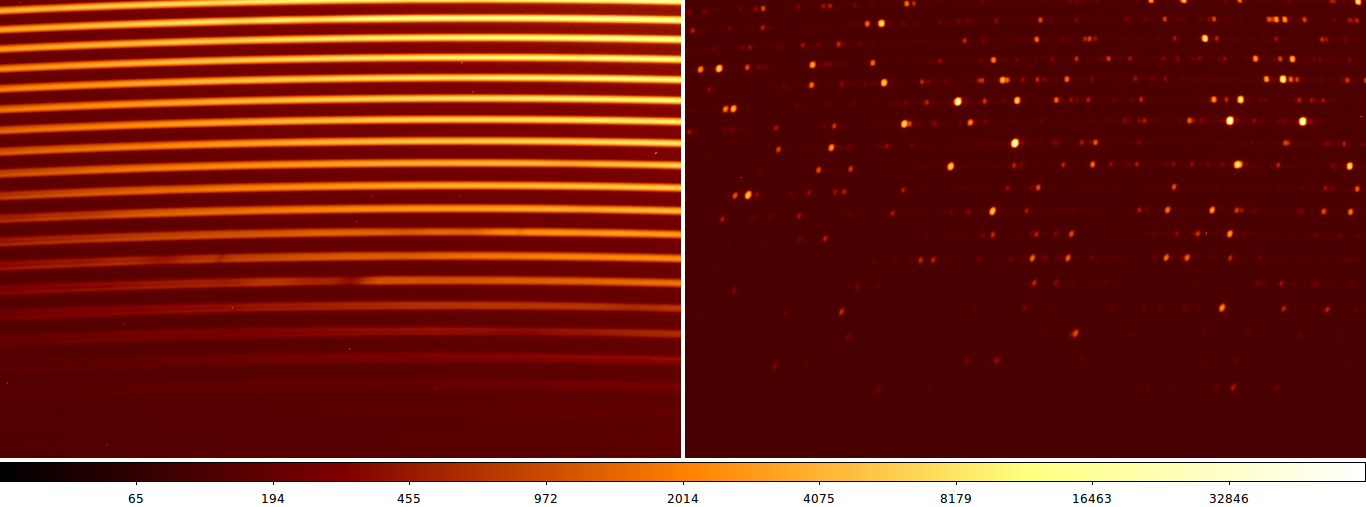}
\caption{Tilted lines in blue part of the spectrum of star beta Lyrae (upper image), zoomed part of lines (left) and straight lines of comparison spectrum (right).}
\label{fig.tilt1}
\end{figure}

\chapter{IRAF}
\label{cha.iraf}
\section{Main Characteristics}
IRAF~(Image Reduction and Analysis Facility) is the system software  project from 80's developed by Peter Shames along with Doug Tody, the chief programmer and the head of IRAF group in the National Optical Astronomy Observatory\footnote{NOAO.} in Tucson, Arizona. This software was designed to reduce and analyze scientific data.\

Its main capabilities include reduction of CCD images and spectral data, the detection of the air masses, the calculation of the Julian heliocentric date and the international coordinate system and more and more other.\

Although, more that 30 years passed since IRAF was first introduced to the scientific community, it is still one of the most popular software used by professional astronomers.\

The latest version IRAF v2.16 was updated on 22. of March 2012 and is freely available for some platforms of GNU/Linux and Mac OSX free of charge. Moreover it includes Virtual Observatory (VO) features like URL access to remote data, search in VO data services by the key words, obtain parameters of chosen object and display its images with \texttt{ds9}\footnote{SAOImage DS9, an application for visualization of astronomical data and displaying images.}, load the table to the VO tool for further use, for instance TOPCAT (Tool for OPerations on Catalogues And Tables) \cite{iraf_noao}.

Native language of IRAF is SPP (Subset Pre-Processor) a portable pre-processor language which resembles C. \cite{iraf}

\section{Installation}
The biggest advantage of the new version is simplification of the installation, which was the problem in previous versions I have experienced on my own, while I~was working on my bachelor thesis\cite{bc}. Back then the installation was difficult in such a~degree, that I~needed to use functioning version on the server at Stellar Physics Department at Astronomical Institute in Ondřejov.\\

The IRAF latest version with the news and updates are available on its websites\footnote{Direct link to version for Linux- 32bit:\

\texttt{htp://iraf.noao.edu/iraf/v216/PCIX/iraf.lnux.x86.tar.gz}}.
Nevertheless, it is important to download X11iraf\footnote{Direct link for Linux- 32bit:\

 \texttt{http://iraf.noao.edu/x11iraf/x11iraf-v2.0BETA-bin.linux.tar.gz} } too\cite{iraf_x11}, for graphical support.\

Once the IRAF 2.16.1 and X11iraf\footnote{This package is installed in $\sim$/iraf/iraf/ within all of the README files and RELEASE notes.} are installed and unpacked, the instruction in \texttt{$\sim$/iraf/iraf/README.install} file is followed. Installation script is dependent on already installed packages, needed for further terminal use of IRAF. The \texttt{tsch} enhanced Unix C shell and utilities for terminal use and X11 libraries the \texttt{libxss1}, \texttt{libncurses5} and \texttt{libXmu6:i386} installed as a~superuser\footnote{\$sudo apt-get install tcsh libxss1 libncurses5 libXmu6:i386}.\

The difference from the older version is that there is an global user login file for IRAF command language stored in \texttt{$\sim$/.iraf/login.cl}. It sets the terminal and image type, editors, packages and more. Convenient is to define often used packages by IRAF reduction such as \texttt{noao} for optical astronomy packages, the image reduction packages \texttt{imred}, \texttt{ccdred} and \texttt{echelle}.\  

Moreover, dynamic external packages created in \texttt{/iraf/iraf/extern/} directory, can be loaded and will be automatically defined when the program is started. Once again, directions in  file is followed. \\

\section{Reduction with IRAF}
IRAF runs with command \texttt{iraf} typed in terminal. Consequently the new window is opened in command line (cl) and ready to fulfill any wishes. Highly recommended for reduction IRAF is the manual by Ch. W. Churchill \cite{church} or by Aoki Wako \cite{redu}. \

\subsection*{The OESPACKAGE pipeline}

Petr Škoda produced partly automatic pipeline for reduction of echelle data from the Ondřejov Echelle Spectrograph. This useful tool called \texttt{oespackage} is available on the web of Stellar Department and need to be included into the \texttt{/iraf/iraf/} directory. The update of this package was needed, since it was created in 2009 and some updates and  arrangements were made on the spectrograph. \

Important is the presence of \texttt{login.cl}, \texttt{coude.dat} to define OES as a default instrument for further analysis with IRAF and to create \texttt{loginuser.cl} with the path to this package:\\

\texttt{set oespath="/iraf/iraf/oespackage/"} \

\texttt{task \$oes = "oespath\$oes.cl"}\

\texttt{keep} \\

To test if it is all set right, simply run IRAF in this directory with data for reduction, login.cl, loginuser.cl and coude.dat: \texttt{$\>$iraf}, which opens command line and than type \texttt{oes}. Subsequently all of the oes tasks (including \texttt{transoes, prepoes, procoes, listoes}, and others) should be displayed. 
\subsection*{Steps of the reduction}
\begin{enumerate}

\item The reduction starts with \texttt{transoes} task for transforming all files according to x-axis and y-axis. This process is also adding ''t" in front of the file name, for instance the object will be renamed into \texttt{tc201307040002.fit}. 

\item The directory \texttt{prepoes} is for trimming already transformed images (starting with tc*) to final forms of required size adjusted in order to remain the same as OPERA is using, to trim its images. Since the \texttt{prepoes} is using IRAF's \texttt{ccdproc} tasks, the size can be defined by simply changing the parameters in \texttt{\$/iraf/iraf/oespackage/} \
\texttt{/prepoes.cl/}.\\

\texttt{ccdproc.trimsec="1:2048,420:1450"}\\

The process is loading the transformed files and allows the previews to check if images are suitable for further process. Different commands offer possibilities to see the head of the file, file information,  ds9 image control. Pressing of key \texttt{ENTER} confirms, that certain image of zero, flat or star is suitable for further analysis.

\item Along with flat and zero combining the task \texttt{procoes} is also used for the bad pixels corrections. Moreover, this changes can be checked by extracting the information about updated files with command \texttt{ccdlist}, for instance, the information about the object will be:\\

\texttt{tc201307040002.fit[2048,50][real][object][][BTZ]},\\

where \texttt{real} says that the image was already processed, followed by the ccdtype of the image and the last are the letters for corrections on the spectrum.\

\item To trace the apertures and extract the data from 2D CCD image it is used \texttt{traceoes}. Concentrating on the update of the oespackage and synchronizing it with OPERA, the reference flat is created by adding the option \texttt{isnewre+} together with \texttt{traceoes}.\

It was important to localize start of the aperture corresponding to those for OPERA. The best option was to concentrate on the position of H$\alpha$ in the object spectrum displayed by \texttt{traceoes}. The process is following: undefined apertures of $\beta$ Lyrae are displayed in interactive window, the aperture containing H$\alpha$ is estimated by trail and error at centered location with value 
approximately 217 and is proved by marking this aperture with \texttt{m}, tracing with \texttt{t} and extracting with \texttt{e} by preview this line should be displayed as shown in Figs. \ref{fig.newre}.\ 

The last picture of this Figure includes final reference flat with defined apertures and how it was the H$\alpha$ found, assigned to the order 44 according to OPERA. Apertures need to be found first by pressing \texttt{f} in iraf interactive window, followed by command \texttt{o 44} for marking the H$\alpha$ aperture, the rest is adjusted by itself starting with 25 and ending with 83. All apertures together are affected after pressing \texttt{a} with meaning ''all". Consequently, the aperture can be adjusted to certain level for all aperture by commands \texttt{:ylevel} , \texttt{a} and \texttt{z}.\

Pressing command \texttt{q}, as is standard for IRAF, saves the procedure, after the fo\-llo\-wing questions will arise as confirmation of tracing, writing apertures of reference flat (\texttt{flatref}) to the database, extracting apertures and reviewing extracted spectra. Moreover \texttt{flatref} is stored in \texttt{/iraf/iraf/oespackage/} and will be used as a reference for star, which is edited interactively whereas the comparison spectrum automatically, as defined in \texttt{traceoes}.\
\begin{figure}
\centering
\includegraphics[scale=5,width=210pt]{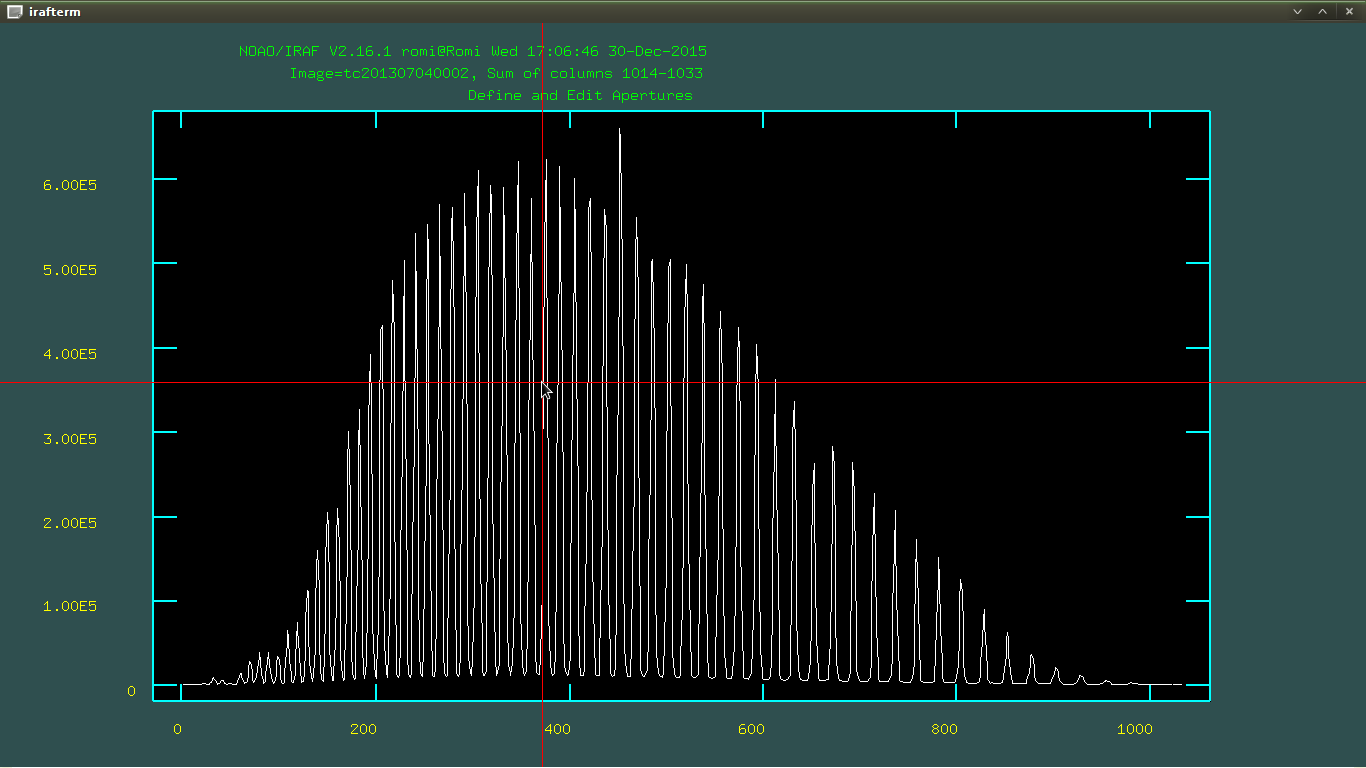}
\includegraphics[scale=5,width=210pt]{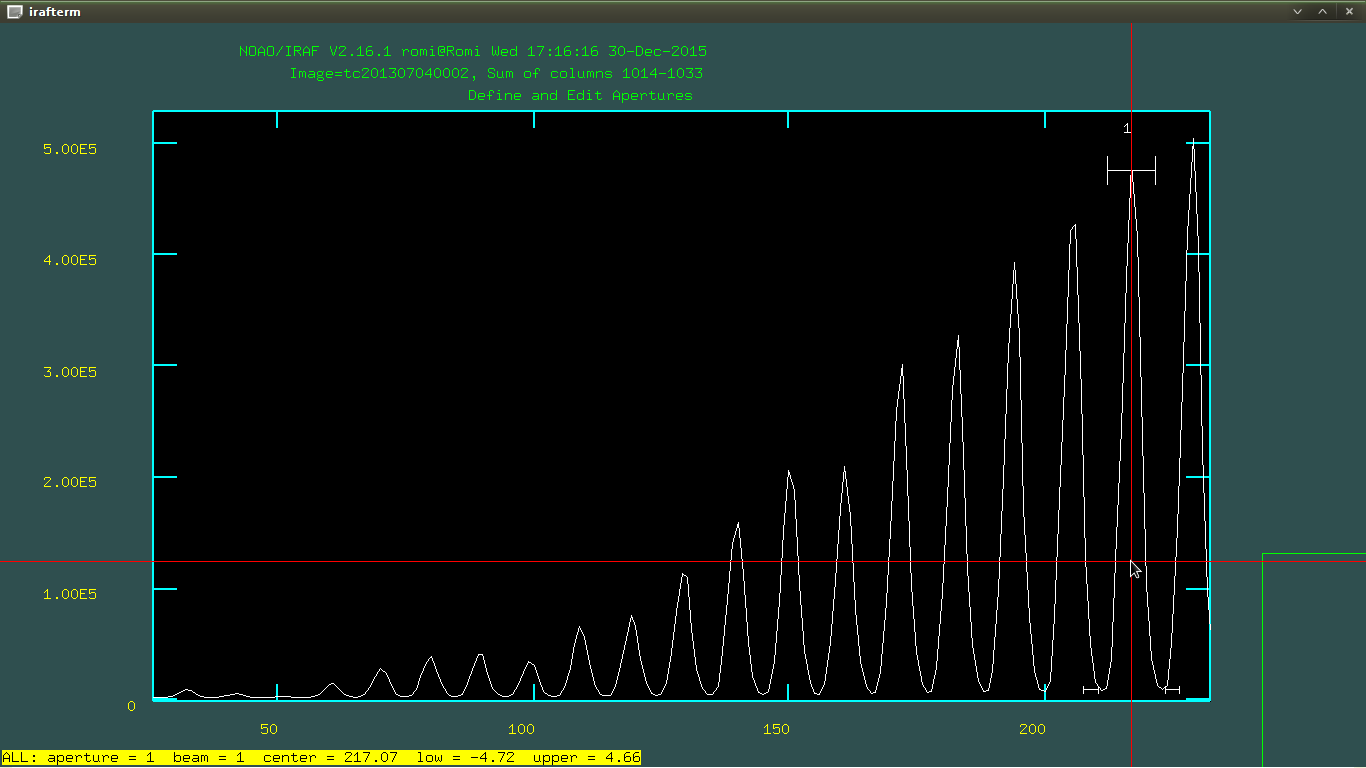}
\includegraphics[scale=5,width=210pt]{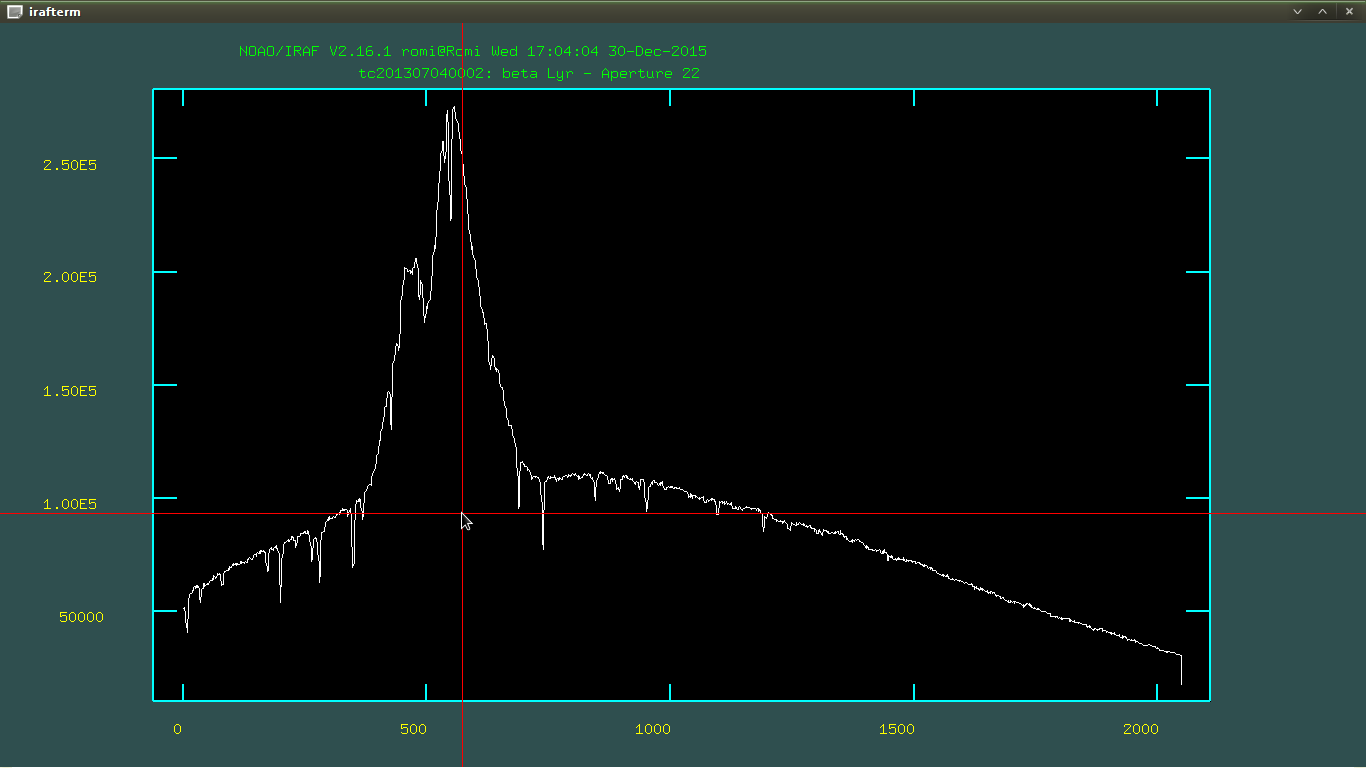}
\includegraphics[scale=5,width=210pt]{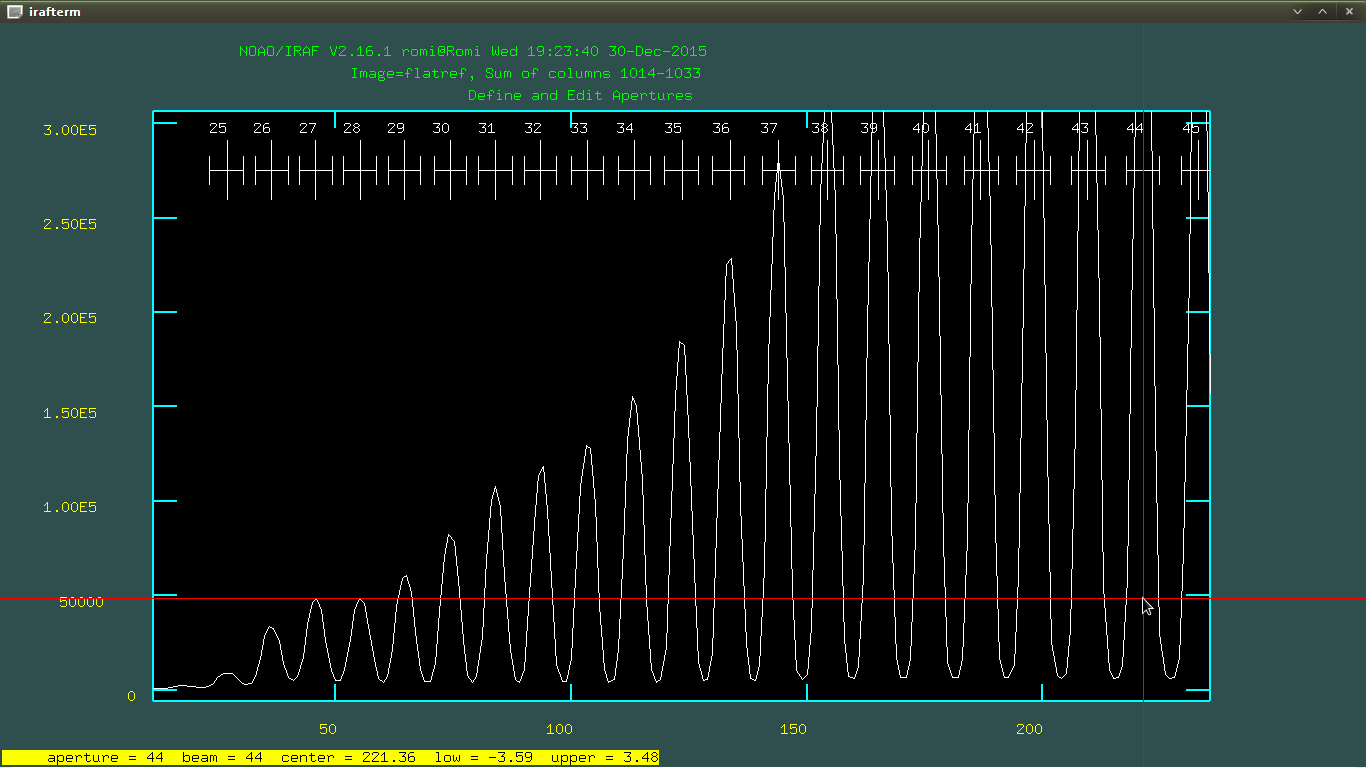}
\caption{Process of defining the numbers of apertures according to those for OPERA.}
\label{fig.newre}
\end{figure}

\item Next \texttt{identoes} is using IRAF's routine for generating a wavelength calibration solution with the list of line wavelengths in calibration lamp exposure, in this case \texttt{comp.ec} stored in \texttt{/iraf/iraf/oespackage/}, the reference comparison set of thorium-argon lines. Lines were identified manually by using prepared identification of spectral lines within orders starting with 37th up to 81th from the OPERA extractions made by Miroslav Šlechta and adjusted for identification by Petr Škoda.

\item Futher follows the command \texttt{ecidentify}, which identifies the wavelength solution and compares it with reference spectrum. This reference spectrum  \texttt{comp.ec} is loaded into the interactive IRAF window by \texttt{:read comp.ec}. Only one order is displayed at the time, key \texttt{k} moves to the next or with keys \texttt{:o 37} to 37th order, in case of sample data the first order with identified lines. Reference lines can be shifted in contrast to present comparison spectrum. To control current state zooming might be useful, performed by moving cursor to the lower left corner of part of the image, that wanted to be examine and using \texttt{w} followed by \texttt{e}  afterwords sliding cursor onto the upper right corner pressing key \texttt{e}. The slight shift appears and is solved by pointing cursor onto the center of line that suppose to be marked and using \texttt{x} for cross correlation of the peaks.\

To determine how accurate
the wavelength solution is enter into the fit mode by hitting the \texttt{f} key. Looking at the RMS residuals to get an idea of
the goodness of fit. As recommended in Churchill's manual \cite{church}, the fitting should start with the lowest orders and slowly increasing it as long as the sine-shape functional dependence is removed. To improve the fit,
points that are obviously incorrect can be deleted by cursor on that point and pressing \texttt{d}. After the residuals are the order of thousandths, leaving the fit mode is preform by \texttt{q} and repeated pressing of this key write the wavelength solution to the database.

\item Last task performed automatically is called \texttt{dispcaloes}. The wavelength solution is applied in order to get dispersion corrected spectrum. This command is using as a reference spectrum \texttt{comp.ec}. \
Important is to define parameter of Ondřejov observatory to the database of all observatories stored in \texttt{/iraf/iraf/noao/lib/}\
\texttt{obsdb.dat/}, with following parameters:\\

	   \texttt{observatory = "ondrejov"}\

       \texttt{name = "Ondrejov Observatory"}\

       \texttt{longitude = 345:12:59.0}\

       \texttt{latitude = 49:54:38.0}\

       \texttt{altitude = 528.}\

       \texttt{timezone = -1}\\

Dispersion corrected spectra are created in following format:\\

\texttt{$>$ wct201307040002.ec.fit} $\Rightarrow$ beta Lyr\

\texttt{$>$ wct201307040003.ec.fit} $\Rightarrow$  comp.

\item Heliocentric Radial Velocity Calculation will be preform. The IRAF's \texttt{rvcorrect} computes radial velocity corrections and is using extracted wavelength calibrated spectrum of the star. To preview and adjust parameters load \texttt{epar rvcor}:\\

\texttt{PACKAGE = astutil}\

   \texttt{TASK = rvcorrect}\

\texttt{(images = ~~~~~~~) List of images containing observation data}\

\texttt{(imupdat=  ~~~~yes) Update image header with corrections?}\\

And run this task with ":go" from "epar window", or just save with ":q" and run with:\

\texttt{cl> rvcorrect}\\

\item Correspondingly, heliocentric correction is applied by task \texttt{dopcor} with output file saved as \texttt{wtc201307040002$\_$hc.ec.fit} and \texttt{redshift = -VHELIO}.

\item Moreover flat-field can be divided from the extracted spectrum. The task \texttt{imarit} is used to divide final star spectrum and extracted flat file (\texttt{Flat.ec}).\

Although it seems inappropriate to wait for the division till the end of the process, there are crucial reasons for operation like that. Theoretically if the spectrum was adjusted by flat-field at the beginning, that would produce intense peaks and make further analyses more difficult, considering the differences of intensity at the edge of each echelle order.

\item Last task \texttt{scombine} is merging multi-order spectra, whereas parameters group and combine can be adjusted \cite{redu}.\\

\texttt{PACKAGE = echelle}\

   \texttt{TASK = scombine}\

\texttt{(group~~=~~~~~~images) Grouping option}\

\texttt{(combine=~~~~~~~~~sum) Type of combine operation}
\end{enumerate}

\section{Final extracted spectra}
\begin{figure}[h!]
\centering
\includegraphics[scale=5,width=200pt]{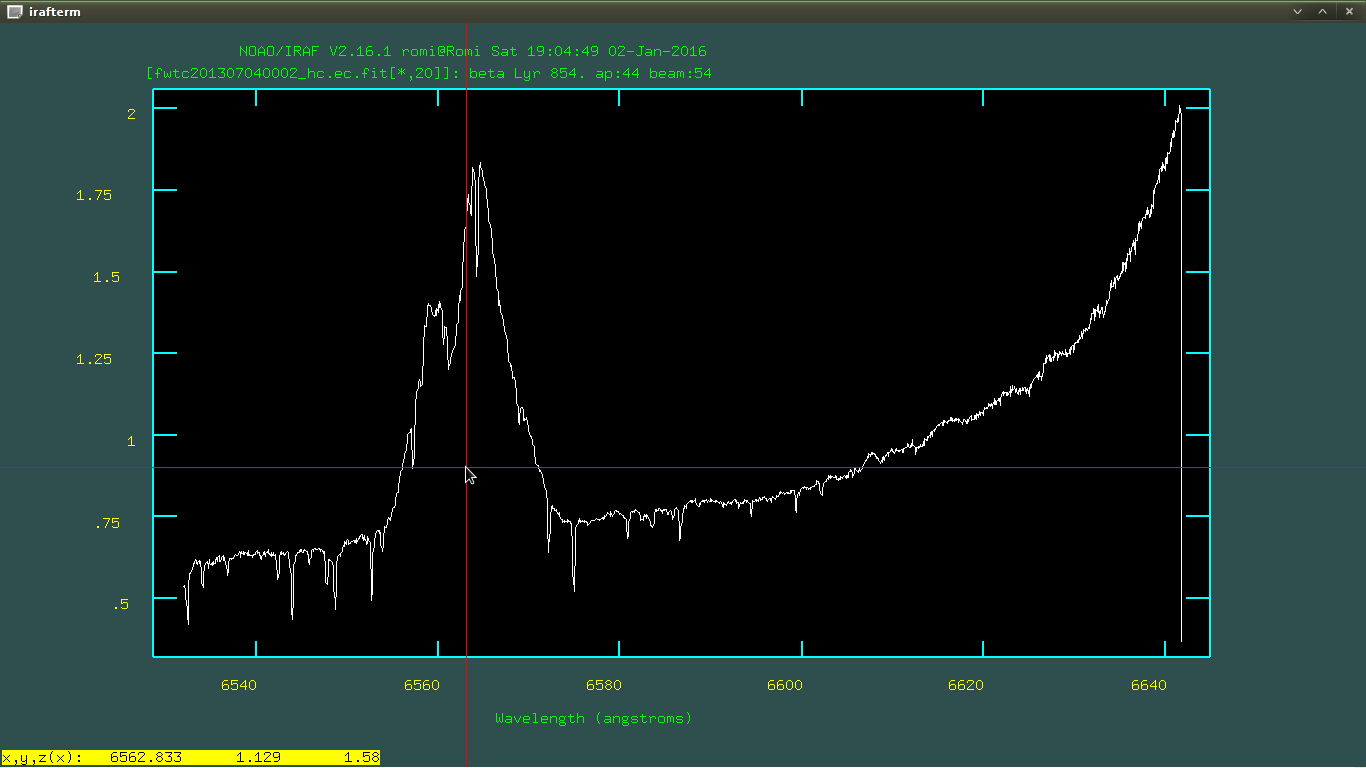}
\includegraphics[scale=5,width=200pt]{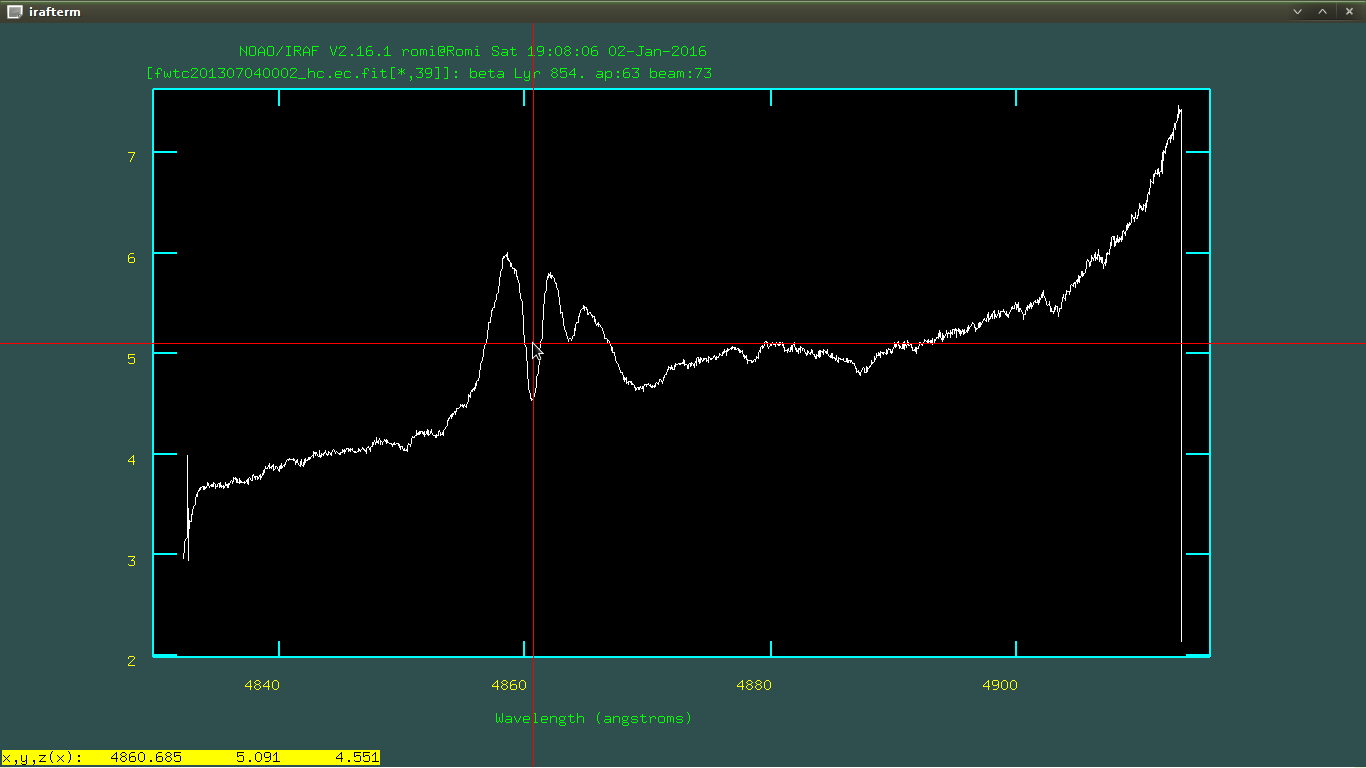}
\includegraphics[scale=5,width=200pt]{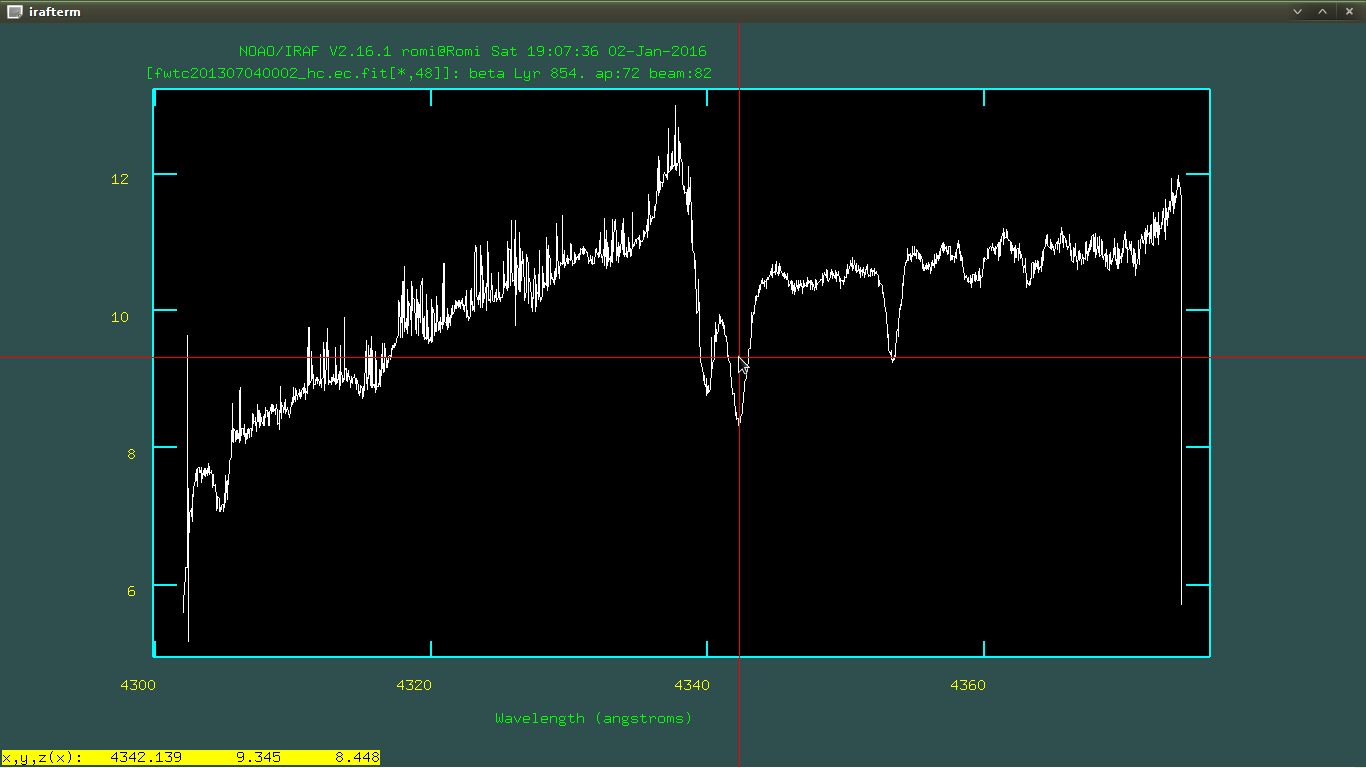}
\includegraphics[scale=5,width=200pt]{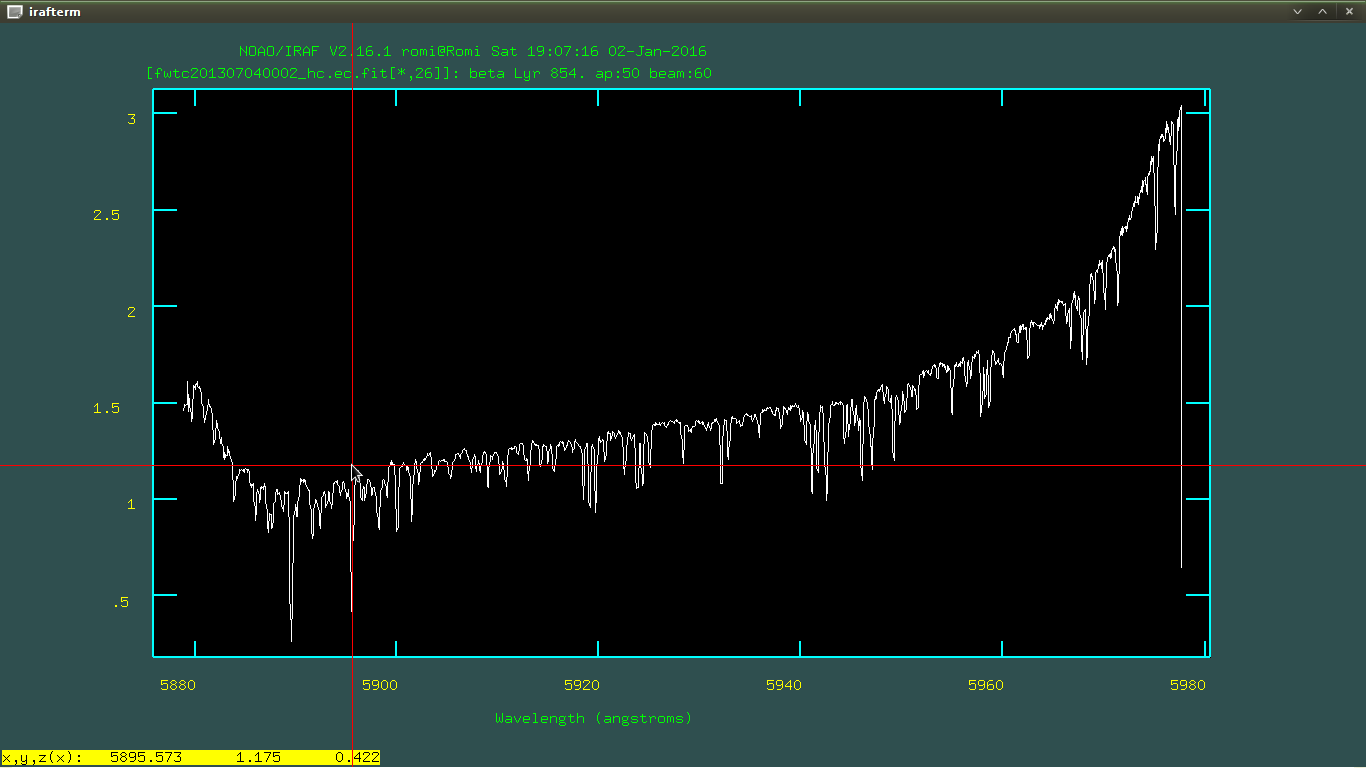}
\caption{Final extracted spectrum for beta Lyrae in different lines with IRAF.}
\label{fig.lines}
\end{figure}

%
Final extracted spectra can be plotted by task \texttt{splot} as demonstrated in Figs. (\ref{fig.newre} and \ref{fig.lines}). Interactive window includes only single order of the spectrum at the time. The move to the next order is accomplished by holding \texttt{SHIFT} key together with pressing left or right parenthesis \texttt{(} or \texttt{)}, respectively. 
\begin{figure}[h!]
\centering
\includegraphics[scale=5,width=450pt]{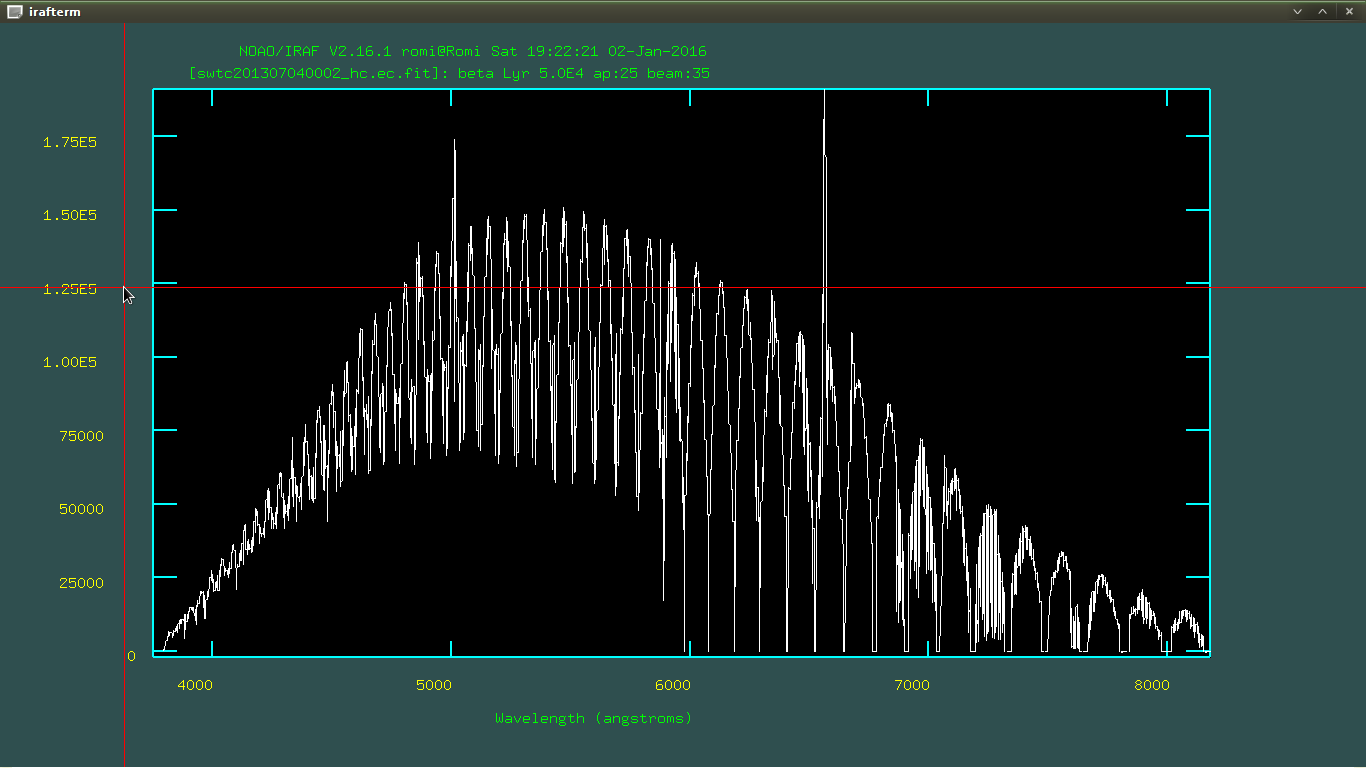}
\caption{Final extracted spectrum for beta Lyrae with IRAF.}
\label{fig.newre}
\end{figure}

\newpage
As a matter of fact, final comparison of spectra from OPERA and IRAF is performed in \texttt{gnuplot}. Therefore, conversion of the products from IRAF in ''.fit" format to ''.dat" files is necessary. Short script was generated for this purpose in Ruby. \\

First, the script in Appendix \ref{sc.ruby1} runs as \texttt{ruby iraf$\_$commands.rb 59 wtc20$\_$} , where number 59 defines the last aperture to be processed and second parameter belongs to the input spectrum, that will be converted.\\

Consequently, it produces set of lines to extract each order separately with \texttt{scopy}, that means one-dimensional files and with \texttt{wspectext} converts 1D files to ASCII header fo\-llo\-wed by data files including two columns with wavelength and intensity.\\ 

\texttt{scopy wtc201307040002.echc.fit[*,1,1] ly1}\

\texttt{wspectext ly1 ly1.dat}\\

The lines created are copied IRAF's command lines. Important is to note, that within IRAF's environment, is not possible to use customized combination ''Copy and Paste", therefore one needs to mark desired lines and to copy them pressing both buttons on mouse together or middle button.\

 After all pairs of commands are successfully done, second script in Appendix \ref{sc.ruby2} will look into current directory named as ''ly*.dat", load each line of files starting with 143 till the end. To run script and save products to final ''.dat" format with \texttt{ruby iraf$\_$filter.rb >bLyr.dat}.
 
This way all files are ready to be examined by \texttt{gnuplot}. But first, the comparison data from OPERA need to be produced in next chapter.

\chapter{OPERA}
\label{cha.opera}
In the field of data reduction and analysis in spectroscopy arises new powerful tool, the software pipeline ''OPERA". OPERA stands for Open source Pipeline for ESPaDOnS Reduction and Analysis and is an Canada-France-Hawaii Telescope(CFHT) open source collaborative software for echelle spectro-polarimetric data. It is fully automatic and able to perform calibration and reduction of data, produce the one-dimensional intensity and polarimetric spectra and much more. The biggest difference from already existing echelle pipelines in Hawaii should be that OPERA is accessible and adjustable to be used for both GNU/Linux and Mac OSX distributions.\

Primary, it is used for data from ESPaDOnS and it is considered to be a replacement for the Upena pipeline using Libre-Esprit software, closed source software developed by Donati in 1995 \cite{donati}. That was the reason for Douglas Teeple and Eder Martioli to de\-ve\-lo\-ped an open source software. Eder Martioli is a~resident astronomer and  part of CFHT Corporation, and Douglas Teeple, his coworker at CFHT, is a~software engineer.\

However, the new statement from CFHT community from October 2015 says, that in 2016 both OPERA and Libre-Espirt will be distributed to the users to get a~feedback, before OPERA will eventually take a~lead.\  

\section{Installation}
OPERA pipeline is available on Sourceforge websites\footnote{Direct link to download opera-1.0 available on\

\texttt{http://sourceforge.net/projects/opera-pipeline/files/opera-1.0/\
opera-1.0-Oct07-2015.tar.gz/download}} \cite{opera_pip}.
Until the current last version from 7th of October 2015 the process of installation was really difficult followed by many errors and incompatibilities, especially for other than MAC distribution. Codes to run the installation were written in bash scripts with dependencies on ESPaDOnS data. In the half of the year 2015 those scripts were transferred to Python by Eder Martioli and also he adjusted them for the OES parameters.\

This last available version is downloaded from the Sourceforge websites. Process of installation was preformed on GNU/Linux Mint (my own notebook) and repeated on Debian (on the server of Stellar Department).\
Several steps toward successful installation need to be followed and will be demonstrated in next paragraphs.\

OPERA package is downloaded and the path to the OPERA is saved in \$OPERA:\\

\texttt{\$ tar xf opera-1.0-Oct07-2015.tar.gz
}\

\texttt{\$ cd opera-1.0/} \

\texttt{\$ OPERA=`pwd`} \\

Very important is to concentrate on the right version of cfitsio library, the C and Fortran subroutines for reading and writing data files in FITS.\

After detailed analysis made by Zdeněk Janák, doctoral student on Masaryk University, it was discovered that last three versions, 3.350, 3.360 and 3.370, are not compatible with OPERA. Problem is in added parameter in those versions, which are slightly different than in OPERA source codes.\ 
The use of 3.340 version is chosen and available on NASA websites \cite{cfitsio}. This package is downloaded, unpacked, configured and installed as following commands demonstrated:\\

\texttt{\$ tar xf cfitsio3340.tar.gz}\

\texttt{\$ cd opera-1.0/} \

\texttt{\$ ./configure --prefix=\$OPERA} \

\texttt{\$ make install} \\

Next the OPERA is configured. Specially for OES, calibration files and pipeline running scripts, where added to OPERA and stored in:\

\texttt{$\sim$/opera-1.0/DefaultCalibration/OES}\ and

 \texttt{$\sim$/opera-1.0/pipeline/OES}. \\

\texttt{\$ cd \$OPERA}\

\texttt{\$ ./configure --prefix=\$OPERA --with-cfitsiolib=lib}\

\texttt{--with-cfitsioinc=include}\

\texttt{\$ make install}\\

\section{Wavelength Calibration}
Important is to notice, that OPERA was designed and adapted for the needs of ESPaDOnS.\
Data in input first wavelength guess files for ESPaDOnS (wcal$\_$ref.dat.gz) are essential for calculation of wavelength solution with OPERA. This set can be obtain by first identifying spectrum from the Oxygen telluric lines then calculating the model terms:
\begin{eqnarray}
\lambda_0(\mathrm{order}) &=& a~+ b/\mathrm{order} \nonumber \\
\lambda_0(\mathrm{order}) &=& a~+ b/\mathrm{order} + c/\mathrm{order}*\mathrm{order} \nonumber, 
\end{eqnarray}
where $\lambda_0$ is the zeroth order term and $\lambda_1$ is the first order term in the pixel-to-wavelength solution. These models calculate $\lambda_0$ and $\lambda_1$ as a~function of order number
for ESPaDOnS. Solution are found in module \texttt{operaWavelengthCalibration.cpp}.\


Although, considering OES the new topic rising up. The linear solution of first guess is not sufficient enough. The new version of OPERA appended an input set of the lines identified manually to the pixel-to-wavelength solution. These lines are used to fit an nonlinear initial solution, which is good enough to detect and identify other lines and converge on a~solution using the current algorithm. \

\subsection*{Manual identification}
The task is clear. Generate file using OPERA script to extract each order separately\footnote{this extraction does not include wavelength calibration} and compare it to the Lovis Pepe Thorium-Argon line atlas for assign to each pixel appropriate wavelength.\

\begin{figure}[h!]
\centering
\includegraphics[scale=5,width=\textwidth]{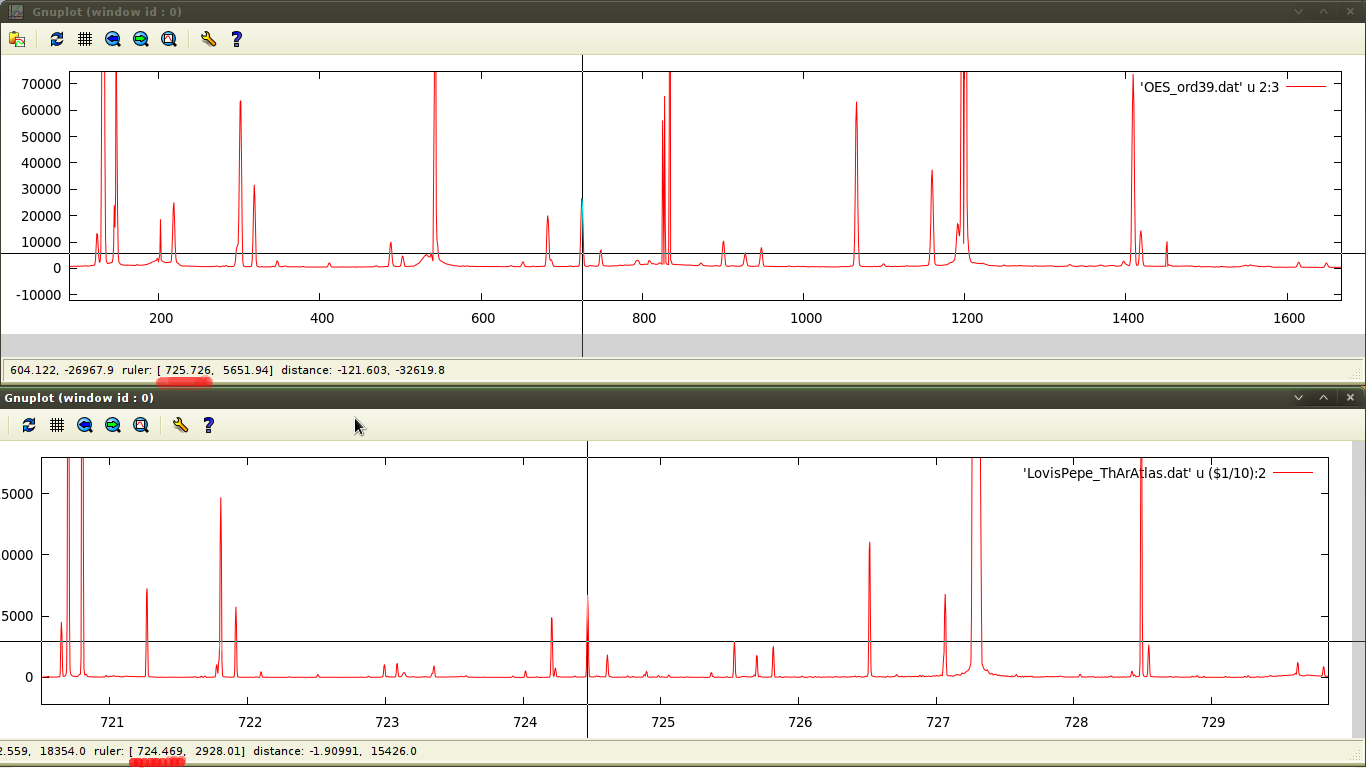}
\caption{Comparison of lines in Lovis Pepe atlas and lines generated for star.}
\label{fig.wcal}
\end{figure}

For the first part is used module \textit{operaWavelengthCalibrationTool} . Files are generated in format for instance ''OES$\_$ord39.dat" and plotted with gnuplot ready to further analyses. New gnuplot window is used to plot certain range corresponding to for instance wavelength range of 39th order in Lovis Pepe atlas. Fig.\ref{fig.wcal} demonstrates the whole process. The top window shows generated lines in 39th order, pixel to intensity dependency, whereas on the window below wavelength dependence of thorium-argon lines is displayed in range 720-730nm .\

Extremely important is to precisely and responsibly identify the same lines in both gnuplot windows. The values of this lines are determined by pointing the cursor on chosen line and pressing "R" letter on keyboard to fix it on position. In red underlined numbers represent the coordinates, pixels in upper window and wavelengths below. This parts of the coordinates are forming data file ''OES$\_$ThArLines.dat". Fig. \ref{fig.thar} demonstrates data file for order 39 composed of three columns: order number, wavelength in nm and distance in pixel. Moreover, created file need to be stored in default calibration directory for OES within opera-1.0\footnote{$\sim$/opera-1.0/DefaultCalibration/OES/}. It is essential for successful extraction of any spectrum of a~star observed by OES.
\begin{figure}[h!]
\centering
\includegraphics[scale=5,width=200pt]{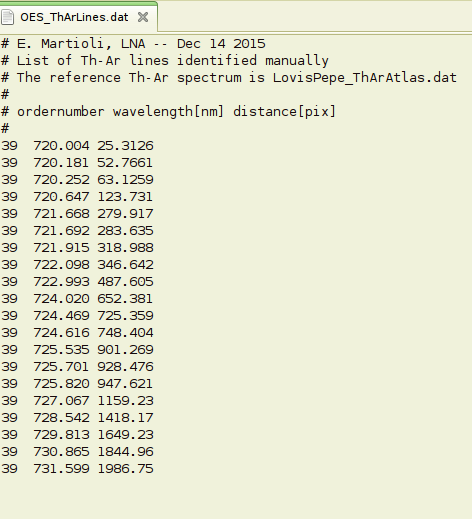}
\label{fig.thar}
\caption{Part of the data file created for wavelength calibration.}
\label{fig.thar}
\end{figure}

\section{Reduction steps}

With produced data file, the final reduction step can be followed. \

Fully-automated OPERA is hiding several calibration steps. Firstly the calibration creates an calibration images, bad-pixel masks, pixel-to-pixel sensitivity map, geometric and instrument profile, aperture and wavelength, to calculate gain and bias followed by reduction for extraction of intensity and polarimetry. Although, OPERA is fully automatic each step of calibration can be performed individually. All of products are in plain-text gzipped format.\

Whereas the reduction includes optimal intensity extraction followed by signal-to-noise calculation, normalization, wavelength telluric correction and creation of final pro\-duct with extension \texttt{*spc.gz}.\\

Data files are the same as were used by IRAF reduction in previous section and stored in created directory:\

\texttt{$\sim$/testOES/data/}.\

The products can be found in\

 $\sim$/testOES/REDUCED/. \\

The reduction is run by the python script \texttt{operaoes.py} and next the paths to directories are set. \\

\texttt{\$ /homedir/opera-1.0/pipeline/OES/operaoes.py }\

\texttt{--datarootdir=/homedir/testOES/data/OES/} \

\texttt{--pipelinehomedir=/homedir/opera-1.0/} \

\texttt{--productrootdir=/homedir/testOES/REDUCED} \

\texttt{--night=2013-07-04} \

\texttt{--product="OBJECTS" -pvt}\\

The last parameter \texttt{--product} defines, what will be produced. Several option are po\-ssi\-ble such as a~''CALIBRATION" or anything from procedures that OPERA is offering: ''GEOMETRY", ''INSTRUMENTPROFILE", ''WAVELENGTH", ''EXTRACT", etc. In all of these cases it will produce all the dependencies before it gets to the final desired product. At last, the options \texttt{-pvt} are set and described the produced plot files \texttt{-p} parameter, for instance \texttt{*eps}, \texttt{*dat} or \texttt{*gnu}, requiered by gnuplot, \texttt{-v} for verbose and \texttt{-t} for printing command lines. Useful testing option is offered too. Including parameter \texttt{-s} runs the simulation of the process. This means it will only print out all command lines on the screen but will not execute anything. It is very useful for testing the steps of calibration and reduction.\\

After successful complete extraction of final spectra, in product directory are created several important files:\\
\begin{enumerate}[label=(\alph*)]

\item Plots for geometry calibration:\

$>$2013-07-04$\_$OESBLADE$\_$100kHz$\_$spcplot.eps\

$>$2013-07-04$\_$OESBLADE$\_$100kHz$\_$geomplot.eps\\

Opera geometry calibration contains detection, tracing, enumerating of spectral orders and the order of center points from master flat-field image. The output can be found in \texttt{*.geom.gz} with the order of polynomial for each order, the polynomial coefficients and coefficient errors.
\begin{figure}[h!]
\centering
\includegraphics[scale=5,width=200pt]{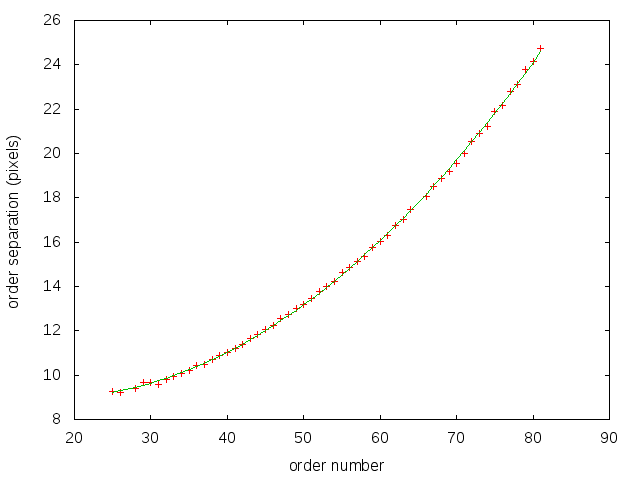}
\includegraphics[scale=5,width=200pt]{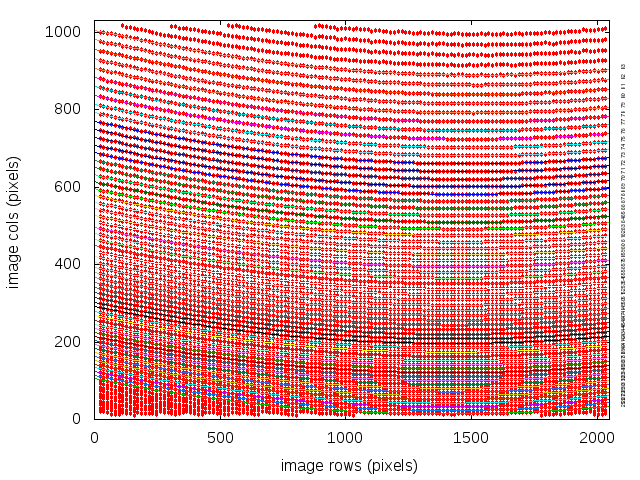}
\caption{Output geometry calibration plot from OPERA.}
\end{figure}

\newpage
\item Plots for the instrument profile calibration:\

$>$2013-07-04$\_$OESBLADE$\_$100kHz$\_$profplot.eps\

$>$2013-07-04$\_$OESBLADE$\_$100kHz$\_$aperplot.eps\\

This part of calibration measures two-dimensional over-sampled normalized instrument profile as a~function of image coordinates at the given size and setting up and calibrating the aperture for extraction with previous module to measure the orientation of a~given slit. The users can select a~number of beams to divide the slit into independent flux measurements as a~result of rectangle shape of this slit. Module also creates two additional smaller apertures on both sides of the slit for background estimates.
\begin{figure}[h!]
\centering
\includegraphics[scale=5,width=200pt]{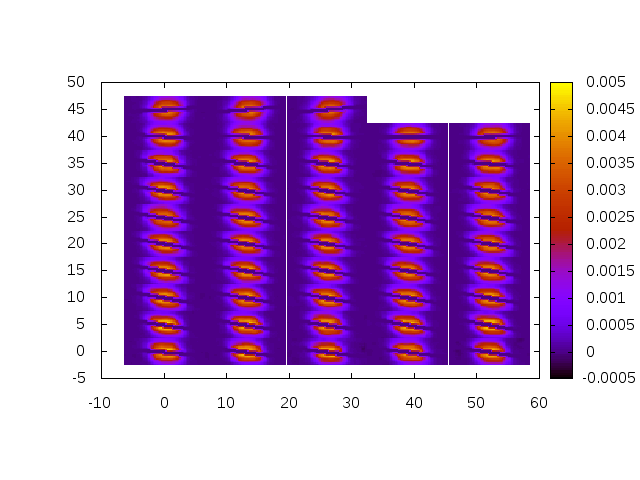}
\includegraphics[scale=5,width=200pt]{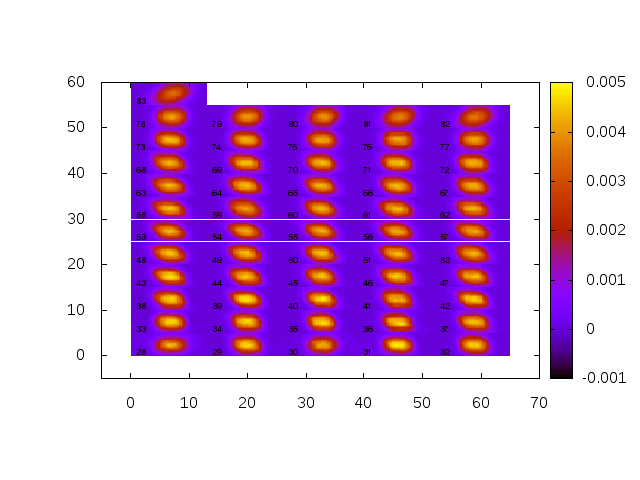}
\caption{Output profile (right) and aperture (left) plot from OPERA.}
\end{figure}
\newpage
\item Wavelength calibration solution plots:\

$>$2013-07-04$\_$OESBLADE$\_$100kHz$\_$waveordsplot.eps\

$>$2013-07-04$\_$OESBLADE$\_$100kHz$\_$wavespecplot.eps\\

Waveleght calibration is used to detect spectral lines and comparing them with the lines of specific lamp. Usually, it is used the Th-Ar atlas. In the beginning it takes a~set of known spectral lines from reference atlas and matches it against lines measured from comparison image as the highest correlation between those lines. Then module stores vector of wavelengths in $\mathrm{nm}$ and distances in $\mathrm{pixels}$ for identified lines and an optimal polynomial is fit to model data
\begin{figure}[h!]
\centering
\includegraphics[scale=5,width=400pt]{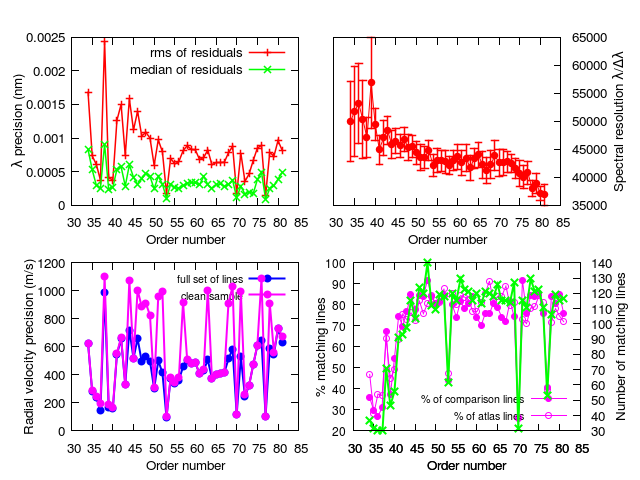}
\caption{Output wavelength solutions plots from OPERA.}
\end{figure}
\begin{figure}[h!]
\centering
\includegraphics[scale=5,width=400pt]{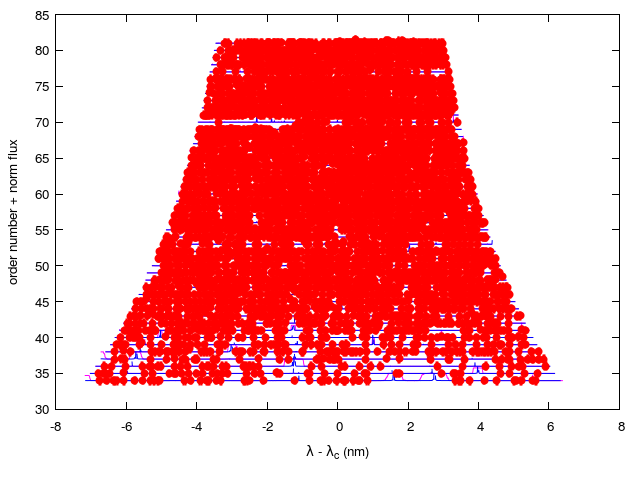}
\caption{Output wavelength solutions plots from OPERA.}
\end{figure}
\newpage
\item The raw extracted spectrum of the comparison lamp:\

$>$2013-07-04$\_$OESBLADE$\_$100kHz$\_$comp.e.gz\\

\item The raw extracted spectrum of the flat field exposure:\

$>$2013-07-04$\_$OESBLADE$\_$100kHz$\_$flat.e.gz\\

\item The raw extracted spectrum of the object:\

$>$c201307040002.e.gz\\
\newpage
\item The reduced spectrum of the object:\

$>$c201307040002.spc.gz\footnote{Note, that all produced files described above are text files with headers including the basic information about the products.}\\

The description inside of the file \texttt{c201307040002.spc.gz} comments the meaning of each column. Interest is put on 6th and 7th column with wavelength heliocentric and telluric corrections, 9th with raw flux, 11th containing normalized flux and 13th for flux of spectrum divided by flat-field. In order to produce whole final spectrum and in H$\alpha$  as demonstrated in Fig.\ref{fig.spectrum} and \ref{fig.halpha}, first it needs to be unpacked with:\

$\$$ \texttt{gunzip c201307040002.spc.gz}

and then following gnuplot script in Appendix \ref{sc.opera_gnu} produces output image.

\begin{figure}[h!]
\centering
\includegraphics[scale=5,width=\textwidth]{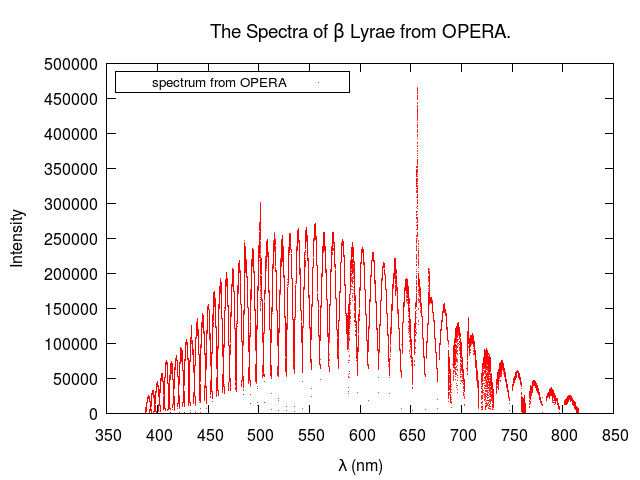}
\caption{Extracted spectrum of the star $\beta$ Lyrae.}
\label{fig.spectrum}
\end{figure}

\begin{figure}[h!]
\centering
\includegraphics[scale=5,width=\textwidth]{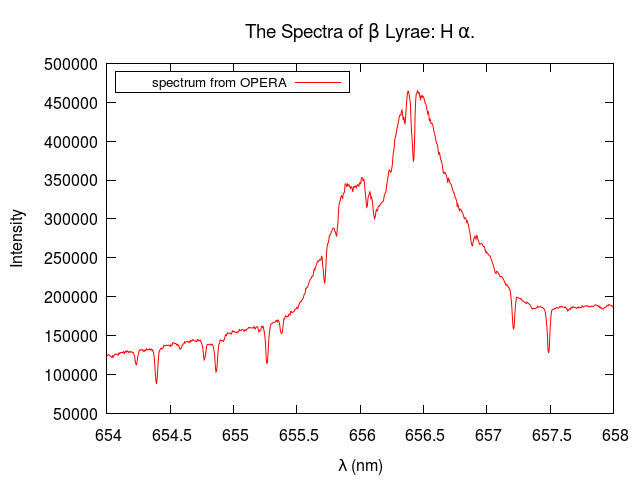}
\caption{Zoom on Halpha in extracted spectrum of $\beta$ Lyrae.}
\label{fig.halpha}
\end{figure}

\end{enumerate}

\newpage
\chapter{Comparison of Final Spectra}
\label{cha.comp}
In this chapter will be performed the comparison of final spectra from OPERA and IRAF. Several gnuplot scripts were written in order to gain higher effectivity and are available in Appendix \ref{opera_com}.\

It should be mentioned, that each spectrum in this chapter was corrected by wavelength telluric and heliocentric correction. The new released version of OPERA from January 2016, provided by Eder Martioli, solves problems with calculation and application of those corrections for data from OES.

\section*{Beta Lyrae}
\label{blyr}
\begin{figure}[h!]
\centering
\includegraphics[scale=5,width=400pt]{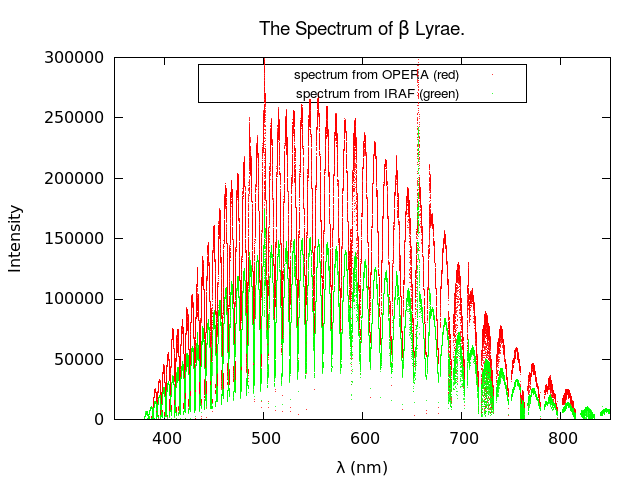}
\caption{Extracted spectrum of the star beta Lyrae.}
\label{fig.bl}
\end{figure}

\begin{figure}[h!]
\centering
\includegraphics[scale=5,width=200pt]{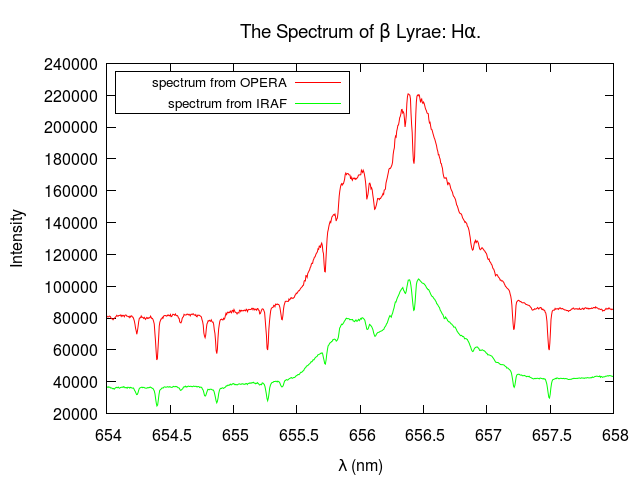}
\includegraphics[scale=5,width=200pt]{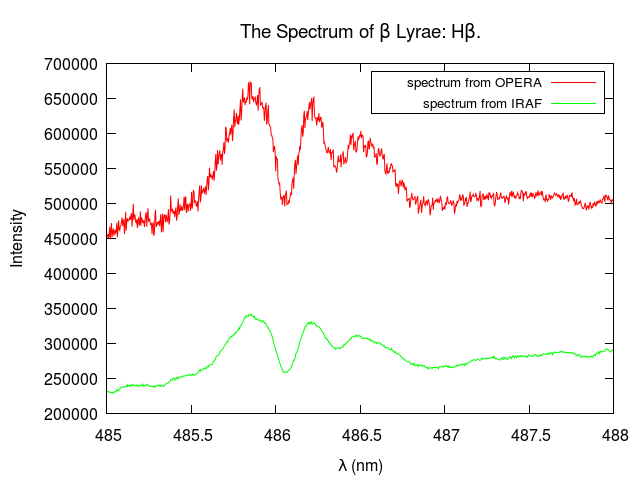}
\includegraphics[scale=5,width=200pt]{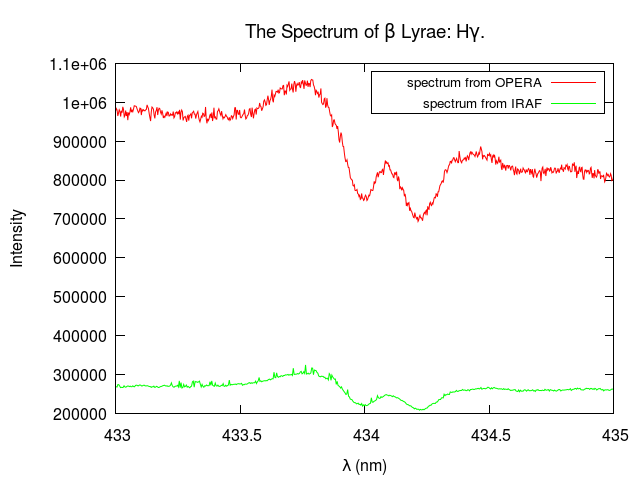}
\includegraphics[scale=5,width=200pt]{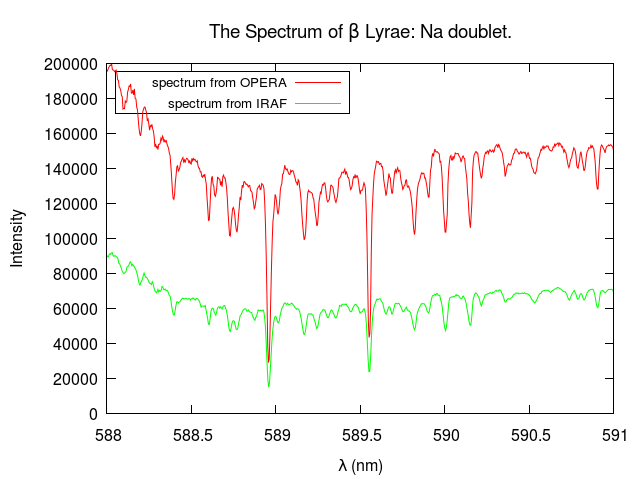}
\caption{Extracted spectrum of the star $\beta$ Lyrae .}
\label{fig.lbl}
\end{figure}

At first the spectra of the star from our sample data, beta Lyrae, are compared in gnuplot. Fig. \ref{fig.bl} and Figs. \ref{fig.lbl} illustrate the whole raw extracted spectrum and flat-field extracted spectra for three lines of Balmer serie and Sodium doublet.

\subsection*{Titled spectral lines}
The problem of the tilted spectral lines for OES (Chapter \ref{cha.data}) unsolvable only for IRAF, has the effects on the near infra-red part of the spectrum for $\beta$ Lyrae. Considering the wavelengths in the range 802-815~nm effects on stellar lines are visible in comparison with stellar lines from OPERA reduction. Such case is depicted in the following Fig. \ref{fig.shift} .\

\begin{figure}[h!]
\centering
\includegraphics[scale=5,width=400pt]{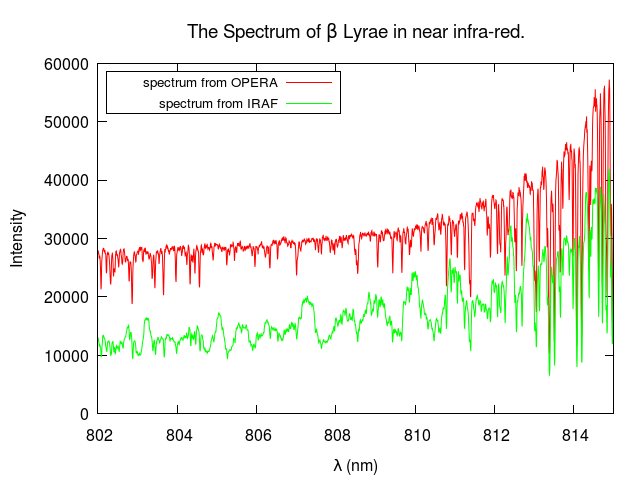}
\includegraphics[scale=5,width=400pt]{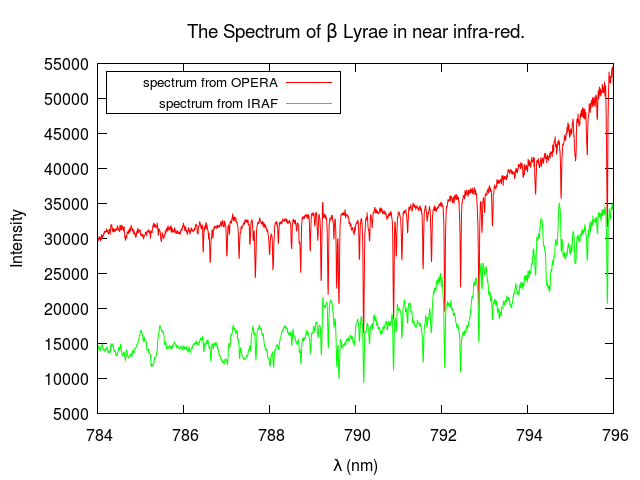}
\caption{Result of tilted spectral line in spectrum of beta Lyrae.}
\label{fig.shift}
\end{figure}

%
%
\section*{Epsilon Aurigae}
\begin{figure}[h!]
\centering
\includegraphics[scale=5,width=400pt]{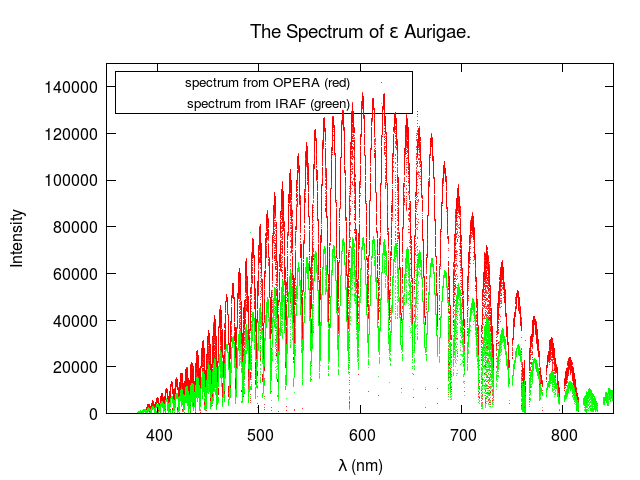}
\caption{Extracted spectrum of the star $\varepsilon$ Aurigae.}
\label{fig.au}
\end{figure}

\begin{figure}[h!]
\centering
\includegraphics[scale=5,width=200pt]{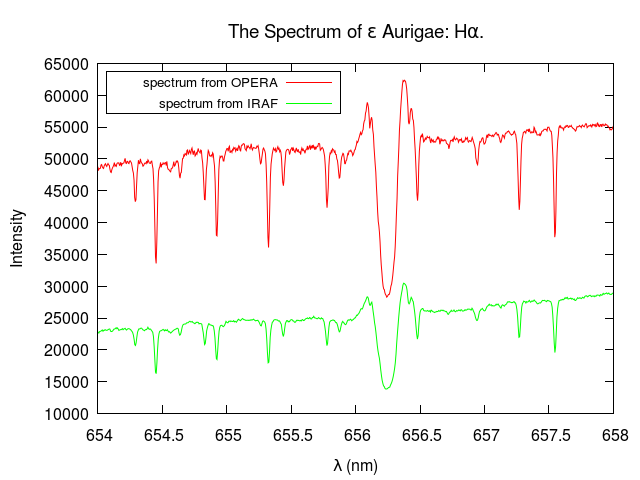}
\includegraphics[scale=5,width=200pt]{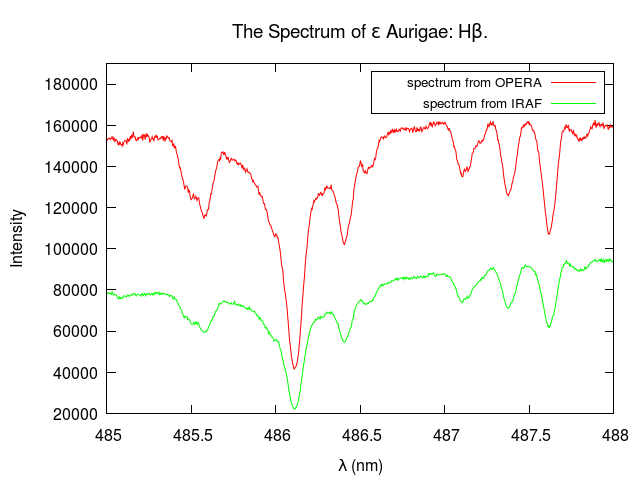}
\includegraphics[scale=5,width=200pt]{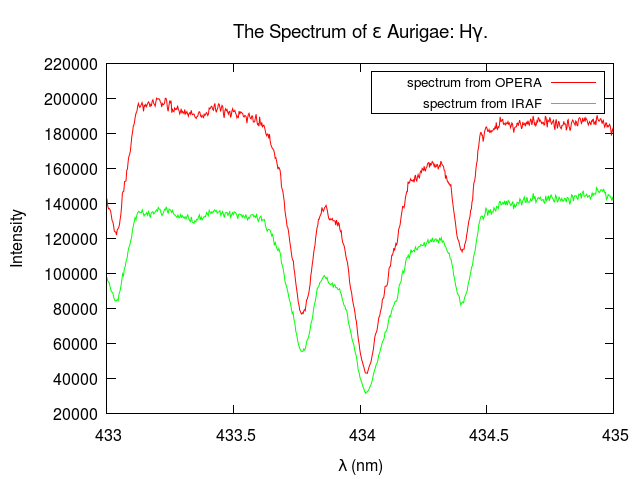}
\includegraphics[scale=5,width=200pt]{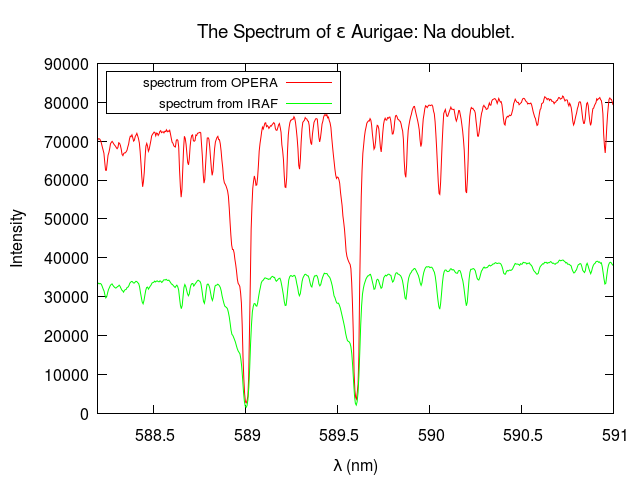}
\caption{Extracted spectrum of the star $\varepsilon$ Aurigae .}
\label{fig.au1}
\end{figure}

Data from OES observed 30th of August 2015 by Miroslav Šlechta from Stellar Department of Astronomical Institute in Ondřejov. The final spectrum of the star epsilon Aurigae, also known as Almaaz, located sightly below Capela in Auriga constellation is displayed in Figs. \ref{fig.au} and \ref{fig.au1}. Comparison of spectra from both pipelines is shown.

\clearpage
\subsection*{Standard versus Echelle spectrograph}
\begin{figure}[h!]
\centering
\includegraphics[scale=5,width=\linewidth]{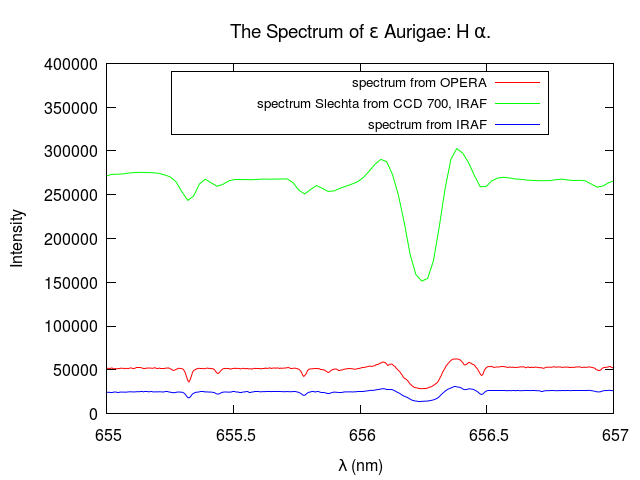}
\caption{Extracted spectrum of the star $\varepsilon$ Aurigae.}
\label{fig.st}
\end{figure}

Miroslav Šlechta observed $\varepsilon$ Aurigae on 30th of August not only with echelle but also standard spectrograph (CCD 700). The results in \texttt{H$\alpha$} of comparison extraction in OPERA and IRAF for echelle and in IRAF for standard are shown in Fig. \ref{fig.st}.

\newpage
\section*{Alpha Canis Minoris}
\begin{figure}[h!]
\centering
\includegraphics[scale=5,width=\linewidth]{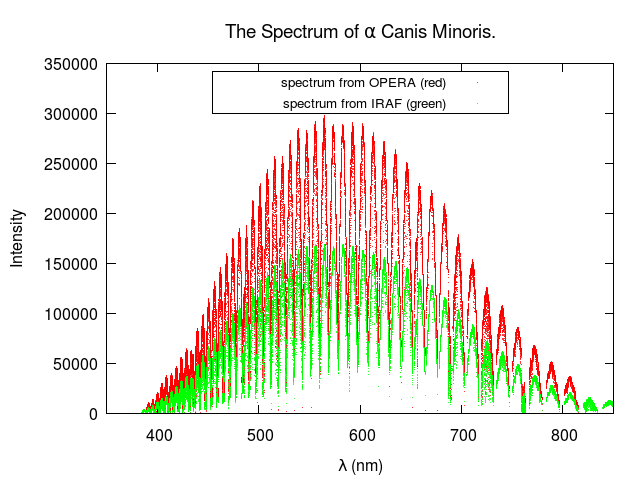}
\caption{Extracted spectrum of the star $\alpha$ Canis Minoris.}
\label{fig.wCmi}
\end{figure}

\begin{figure}[h!]
\centering
\includegraphics[scale=5,width=200pt]{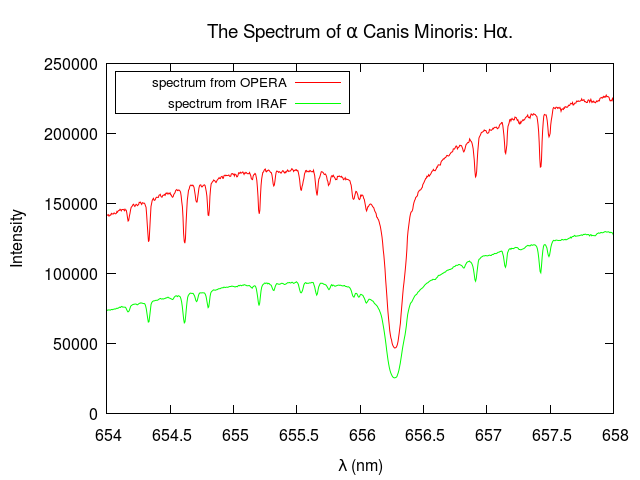}
\includegraphics[scale=5,width=200pt]{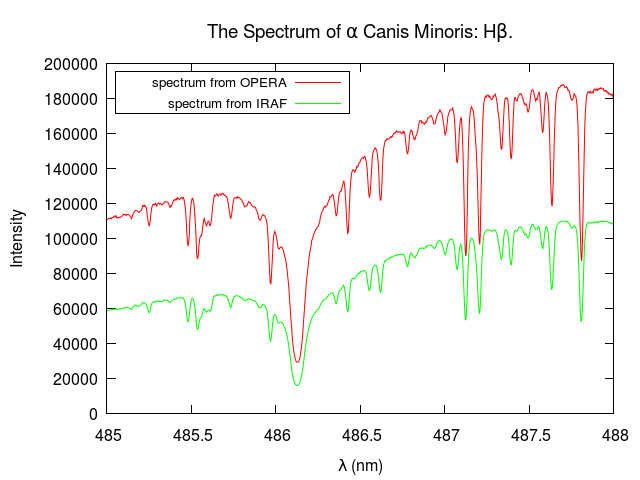}
\includegraphics[scale=5,width=200pt]{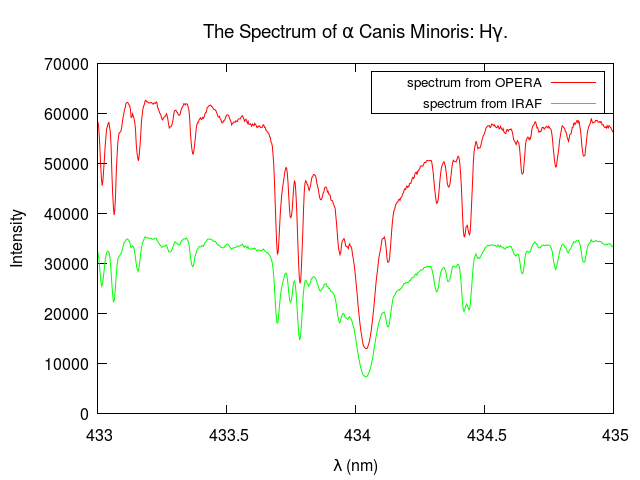}
\includegraphics[scale=5,width=200pt]{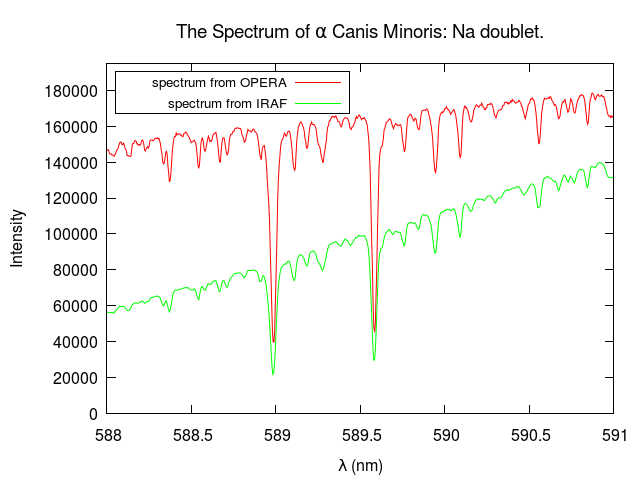}
\caption{Extracted spectrum of the star $\alpha$ Canis Minoris .}
\label{fig.cm}
\end{figure}

In April 2015, alpha Canis Minoris, Procyon was observed. The whole spectrum together with Na and H lines of this F-type star are plotted in Figs. \ref{fig.wCmi} and \ref{fig.cm}. 
\clearpage
\subsection*{Overlapping orders}
\begin{figure}[h!]
\centering
\includegraphics[scale=5,width=400pt]{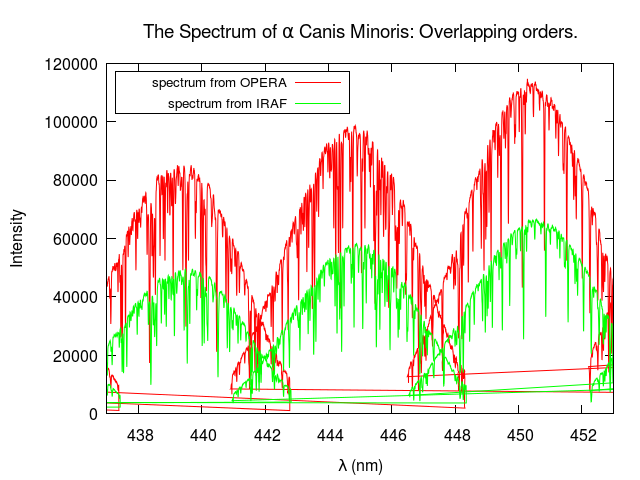}
\caption{Overlapping orders of spectrum of the star $\alpha$ Canis Minoris.}
\label{fig.over}
\end{figure}

\begin{figure}[h!]
\centering
\includegraphics[scale=5,width=400pt]{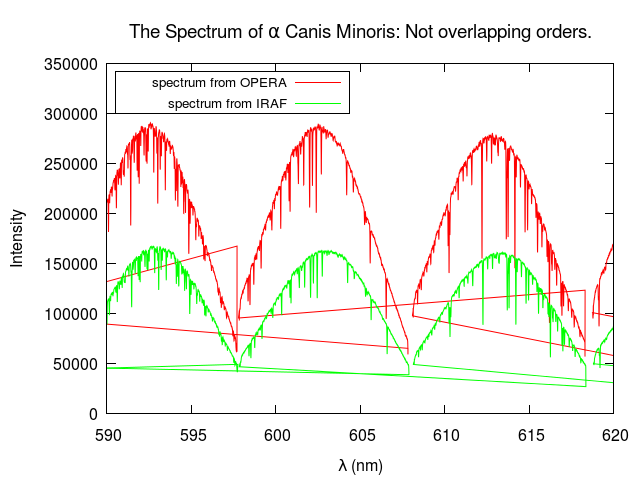}
\caption{Orders do not overlap at the wavelength 600 nm  anymore.}
\label{fig.over1}
\end{figure}

An example of of overlapping orders around blue part of spectrum gives Fig. \ref{fig.over}. On the other hand, in OES echelle spectrum from 600~nm orders start to be separated. This is illustrated in Fig. \ref{fig.over1}. 

\clearpage
\section*{Eta Ursae Majoris}
\begin{figure}[h!]
\centering
\includegraphics[scale=5,width=200pt]{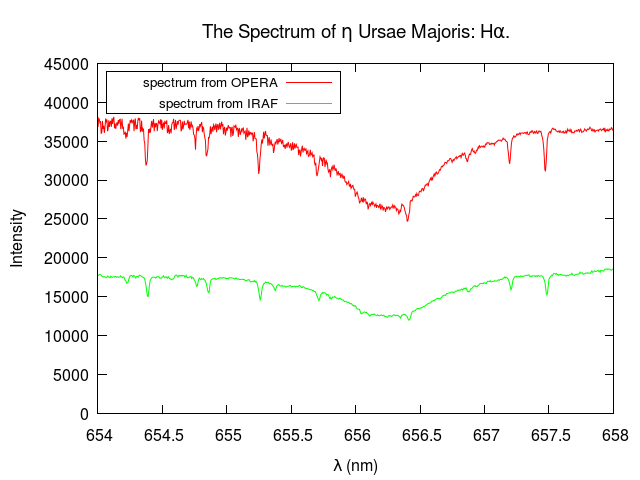}
\includegraphics[scale=5,width=200pt]{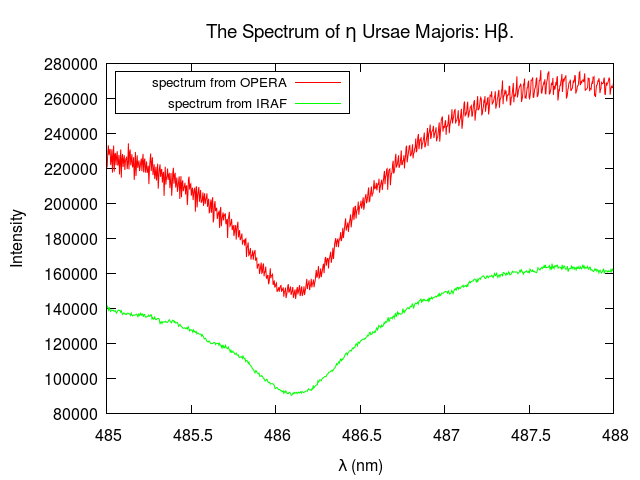}
\caption{Extracted spectrum of the star $\eta$ Ursae Majoris.}
\label{fig.saur}
\end{figure}

The same night were exposed much more stars than only $\alpha$~CMi. Observations of the most eastern star of constellation Ursae Majoris, $\eta$~UMa took a place during this night too. Traditionally named Alkaid, it is more hotter star than $\alpha$~CMi, what corresponds to the width of its lines plotted in Fig. \ref{fig.saur}.
 
\begin{figure}[h!]
\centering
\includegraphics[scale=5,width=400pt]{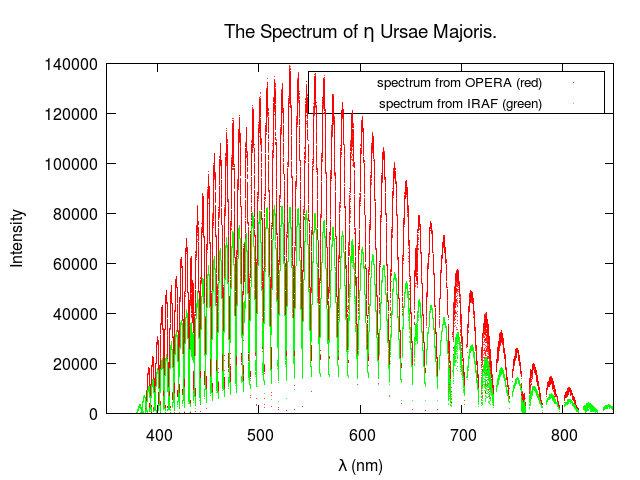}
\caption{Extracted spectrum of the star $\eta$ Ursae Majoris.}
\label{fig.eUm}
\end{figure}

%
\section*{Spectra with Iodine cell}
\begin{figure}[h!]
\centering
\includegraphics[scale=5,width=400pt]{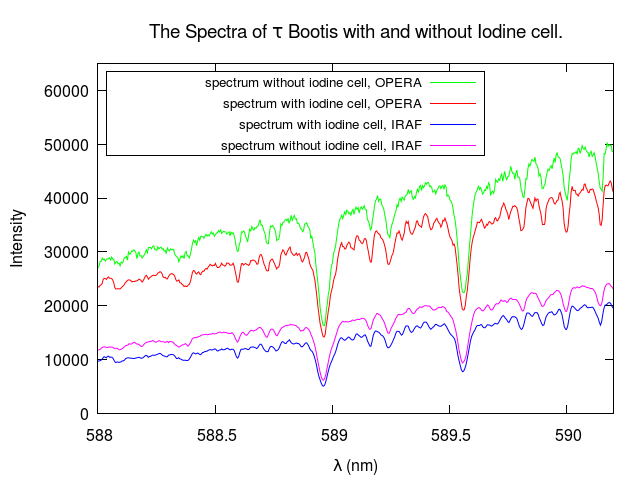}
\caption{Extracted spectrum with and without Iodine cell.}
\label{fig.opir}
\end{figure}

\begin{figure}[h!]
\centering
\includegraphics[scale=5,width=400pt]{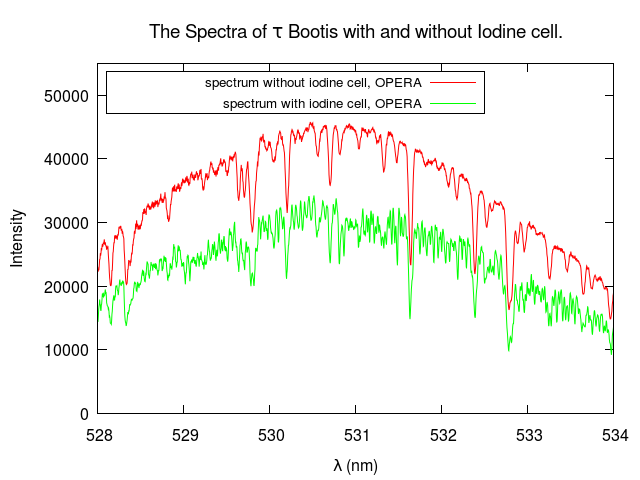}
\caption{Extracted spectrum with and without Iodine cell zoomed in green part.}
\label{fig.ic_2}
\end{figure}

\begin{figure}[h!]
\centering
\includegraphics[scale=5,width=400pt]{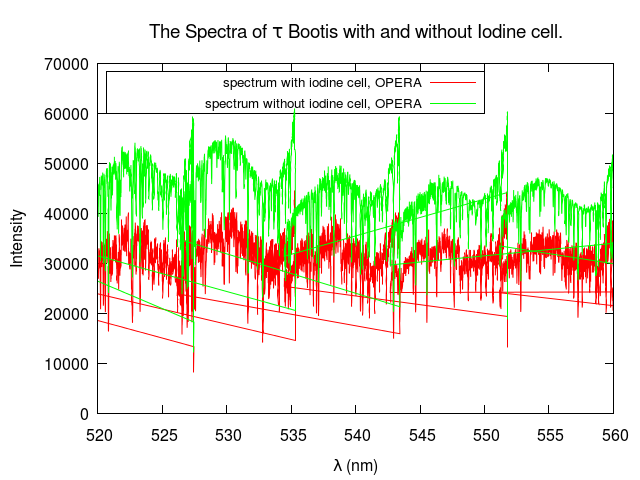}
\caption{Extracted spectrum with and without Iodine cell in range of the highest absorption of Iodine cell.}
\label{fig.ic_1}
\end{figure}

As it was noted in the Chapter \ref{iodine}, Ondřejov echelle spectrograph has an Iodine absorption cell. Sample data observed with and without Iodine cell are presented in this section.  As is shown in Figs. (\ref{fig.opir}, \ref{fig.ic_2} and \ref{fig.ic_1}) spectra exposed with iodine cell contain considerably more lines, evoke the feeling of spectrum suffering from noisy signal. However, as mentioned before it is just dense forest of absorption lines in range between 4800~\AA ~and 6100~\AA ~. 

\section*{Spectrum of stars in H$\alpha$}

\begin{figure}[h!]
\centering
\includegraphics[scale=5,width=400pt]{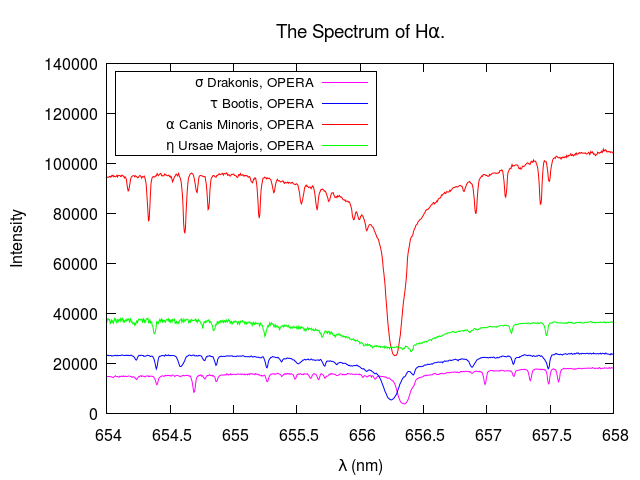}
\caption{H$\alpha$ in spectrum of different stars from April 2015.}
\label{fig.ha}
\end{figure}

Last spectrum presents spectra of different stars  in astronomically important line, Hydrogen, from the 9th April 2015. Spectrum was calibrated that night for $\sigma$~Drakonis, $\tau$~Bootis, $\alpha$~Canis Minoris and $\eta$ Ursae Majoris. In the Fig. \ref{fig.ha} the hottest star, spectral type B3V, of this sample is represented by broad H$\alpha$ line. Because of the fact that this line has different value of radial velocity for each star, the lines are not fitting on each other.  
\chapter{Summary}
%
%
In this thesis was the analysis of different reduction tools in spectroscopy conducted. The software was produced in 80s, but still often used by wide field of astronomical community. Even such a popular tool called the Image Reduction and Analysis Facility, IRAF, has its weaknesses. In other words, it is suitable for any kinds of reduction a analysis of astronomical data, except data from OES,  Ondřejov Echelle spectrograph. The main reason is the problem of tilted  spectral lines with respect to echelle orders widely present in spectra of OES. Because IRAF does not include any tool dealing with this problem, it is not the perfect choice for data analysis of echelle spectra. \

On the other hand, new open-source software was established and managed by Eder Martioli, called OPERA for an Open-source Pipeline for ESPaDOnS Reduction and Ana\-ly\-sis. The biggest inspiration of adapting this automatic pipeline for OES is, that within lots of useful modules and scripts for automatic reduction, extraction and different analysis of spectrum, it contains modules for straighten up tilted spectral lines. \

Thus the results in Fig.\ref{fig.shift} are compatible with theory. This effect is more noticeable in near infra-red part of the spectrum. Although, that is only one advantage of this new tool. OPERA is in comparison with IRAF, let's say, more user friendly, considerably faster and more efficient in automatic creation of master calibration images, determination of statistical quantities for wavelength calibration as spectral resolution, number of matching lines, wavelength and radial velocity precision and calculation of heliocentric correction, radial velocity corrections. Moreover, the source codes and modules are written in more modern and accessible codes, more easy to understand. All modules are written in \texttt{C}, whereas the running codes in \texttt{Python}. On the other hand IRAF's native programming language is Subset Pre-Processor.


The most important and also the most difficult part of this thesis and process of pro\-du\-cing pipeline for OES is to create wavelength calibration template. It seems like easy task, but sorts of problems convinced us otherwise. Mainly, the principal of first wavelength solution was missing some parameters and was not producing right results. Eder Martioli, after many consultations, offered solution including manual identification of lines of the star computed with Lovis-Pepe Thorium-Argon atlas. Moreover, he added some adjustments to OPERA in order to take this solution as sufficient enough, for calculation of  wavelength calibration. This method was tested on sample data from Petr Škoda and is fully described in Chapter \ref{cha.opera} together with main information and standard reduction and extraction procedures. \

To verify this procedures, OPERA reduction is compared with well know reduction from IRAF. Final spectra from both tools are presented in Chapter \ref{cha.comp}. This chapter also includes spectra from tau Bootis observed with Iodine cell. Results shown in Figs. (\ref{fig.io} and \ref{fig.ic_2}) agree with theory which assumes the decrease of intensity in spectrum observed with Iodine cell. Short description of Iodine absorption cell is introduced within the section \ref{cha.oes} including the main characteristics, scheme and parameters of OES. The whole Chapter \ref{eche} is dealing with an comparison of echelle spectrographs from each part of the world. Closer look is taken  not only on OES, but on OPERA's home-base echelle spectrograph ESPaDOnS, too. The main differences between those two instruments are in the spectral resolution and in the transport of light from telescope to the spectrograph. The fact that the spectra resolution is much higher in case of ESPaDOnS is closely related to the second difference. The light from telescope is fiber-fed to the ESPaDOnS. For OES light travels through slit and by several mirrors fed to the spectroscope. Fiber for OES in coude focus would need to be around 20~m long, together with lower high resolution than for ESPaDOnS, the losses of signal would be more significant. Anyway, fiber-fed OES is highly discussed in Stellar Department. Hopefully, in near future more possibilites of different optical fibers will offer solution. \

Essential application of high-resolution spectrographs is included in this thesis in the Chapter \ref{cha.exo}. More accurate determination of the position and shapes of spectral lines is offering more precise measurements of radial velocity closely related to one of the leading topics in modern Astronomy, search of extra-solar planets. The main characteristics and detection methods are described.\

%
%
Clearly, further research will need to be concentrated on normalization of spectra in OPERA. Two difficulties are raising up. The first is related to properties of star exposed on OES. The astronomical research of Stellar Department is mostly concentrated on hot stars, whose lines can be broad (from rotation too) in such a degree, that they are considered to be part of the continuum. On the other hand, some standard star need to be observed in order to perform flux calibration. Moreover, the normalization of spectrum from IRAF did not show the best result, because of uncalibrated flux from OPERA, this topic was not included into this thesis. Doubtlessly, normalization is important for further analysis and better comparison of changes in spectral lines.

%
%
%
%

\newpage
\appendix

\chapter{Gnuplot Scripts}
\label{cha.gnuplot}
Several gnuplot scripts were written in order to preserve the structure of generated output plots and to make the work more effective, too.\

\section*{Script for histogram of detection methods}
\label{hist}
\begin{lstlisting}
#Gnuplot script producing histogram of detection methods.
reset
unset key
set terminal png font "Helvetica"
set output 'histogram_exop.png'
set xtics rotate out
set style data histogram
set style fill solid border
set style histogram rowstacked
set boxwidth 0.5 relative
set key inside left top vertical Left reverse enhanced autotitles nobox spacing 0.7
set title "Yearly exoplanetary discoveries by using detection methodes."
set xlabel "Year"
set ylabel "Number of discoveries"
set object 1 rect from graph 0, 0 to graph 0.5
set multiplot
plot for [COL=2:7] 'discoveries_methode.txt' using COL:xticlabels(1) title columnheader
unset object 1  
set size 0.56,0.53
set origin 0.11, 0.18
set xtic
set xtic font "Helvetica,8"
set ytic
set ytic font "Helvetica,4"
set boxwidth 0.6 relative
set label "The exclusion of the transit \n methode." at graph 0.4,0.9 center font "Helvetica, 10"
unset xlabel
unset ylabel
#set label "Without transit methode" front
#show label
unset title
set ytic font "Helvetica,9"
set key left center vertical Left reverse enhanced autotitle nobox spacing 0.5
set key font "Helvetica,9"
plot for [COL=2:6] 'discoveries_methode_wt.txt' using COL:xticlabels(1) title columnheader
set label "y=x" at 1,2
\end{lstlisting}
\section*{Scripts for OPERA outputs}
\label{sc.opera_gnu}
For instance script \texttt{opera$\_$bLyr.gnu} will produce  output  image of extracted spectrum from OPERA. 
\begin{lstlisting}
#Script opera_bLyr.gnu generates plot of extracted spectrum.

reset
unset key

set encoding utf8
set terminal pngcairo enhanced font 'Helvetica,12'
set output "opera_bLyr.png"
set title "The Spectra of {/Symbol b} Lyrae from OPERA." font "Helvetica, 14"
set xlabel "{/Symbol l} (nm)"
set ylabel "Intensity"
set key top box 7
plot "c201307040002.spc" u ($6+$7):9 w dots
\end{lstlisting}

\newpage
\section*{Scripts for comparison}
\label{opera_com}
As a sample is chosen gnuplot script for \texttt{spectrum$\_$bLyr.gnu}, which has an output the whole spectrum of $\beta$ Lyrae and has following form:\\
\begin{lstlisting}
#Gnuplot script "spectrum_bLyr" to generate output spectra

reset
unset key

set encoding utf8
set terminal pngcairo enhanced font 'Helvetica,12'
set output "spectrum_bLyr.png" 
set xrange[350:850]
set title "The Spectrum of {/Symbol b} Lyrae." font "Helvetica, 14"
set xlabel "{/Symbol l} (nm)"
set ylabel "Intensity"
set key top center box 7 
set key font "Helvetica,10"
plot "c201307040002.spc" u ($6+$7):9 t "spectrum from OPERA (red)" w d, "betaLyr.dat" u ($1/10):2 t "spectrum from IRAF (green)" w d 
\end{lstlisting}
\subsection*{Script for H$\alpha$}
In case of Balmer serie of hydrogen, just one sample script for H$\alpha$ is introduced. \ \noindent
\begin{lstlisting}
#Script "spectrumHa_bLyr.gnu" generates spectra in H alpha.
reset
unset key

set encoding utf8
set terminal pngcairo enhanced font 'Helvetica,12'
set output "spectrumHa_bLyr.png"
set xrange[654:658] 
set title "The Spectrum of {/Symbol b} Lyrae: H{/Symbol a}." font "Helvetica, 14"
set xlabel "{/Symbol l} (nm)"
set ylabel "Intensity"
set key top box 7 
set key left top font "Helvetica,10"
plot "c201307040002.spc" u ($6+$7):9 t "spectrum from OPERA" w l lt 1 lw 1 , "betaLyr.dat" u ($1/10):2 t "spectrum from IRAF" w l
\end{lstlisting}
Scrips for rest of the lines differs only in settings of output, xrange and  as following:\\ \noindent
\begin{lstlisting}
set output "spectrumHb_bLyr.png"
set xrange[485:488] 
set title "The Spectrum of {/Symbol b} Lyrae: H{/Symbol b}." font "Helvetica, 14"
\end{lstlisting}
Consequently, files after \texttt{plot} includes names of files from both OPERA and IRAF extractions, in order to compare them visually. \noindent

\chapter{Ruby Scripts}
\label{sc.ruby}
\section*{Ruby for 1D to ASCII}
\label{sc.ruby1}
Ruby scripts used for converting ''.fit" format to ''.dat" format. First script produces set of lines, one to extract each order separately with \texttt{scopy}, that means one-dimensional files and with \texttt{wspectext} converts 1D files to ASCII format.  
\begin{lstlisting}
# iraf_commands.rb
(1..ARGV[0].to_i).each do |n|
  puts "scopy #{ARGV[1]}.fit[*,#{n},1] ly#{n}"
  puts "wspectext ly#{n} ly#{n}.dat"
end
\end{lstlisting} 
\section*{Ruby for final data file}
\label{sc.ruby2}
Second simply looks at the each file with ''ly*.dat" in current directory and extract lines of this file starting with 143 till the end. 
\begin{lstlisting}
#iraf$\_$filter.rb
Dir['ly*.dat'].each do |file|
  lines = File.readlines(file)

  puts lines[142..-1]
end
\end{lstlisting}

\chapter{Manuals}
Last appendix contains printed versions of reduction summary for both OPERA and IRAF reduction. Those manuals, from Wikipedia portal of Stellar Department\footnote{\texttt{https://stelweb.asu.cas.cz/wiki/index.php/OPERA}}.\

To this thesis is also attached DVD containing IRAF and OPERA manuals for installation and reduction, full set of final test products from OPERA, reduced data from OPERA and IRAF together with gnuplot scripts to generate input images and their comparison. 

%

\end{document}